\documentclass{aa}  

\usepackage{graphicx}
\usepackage{amsmath}
\usepackage{txfonts}
\def\adaptahop    {{\sc adaptahop}}
\def\music    {{\sc music}}
\def\ramses    {{\sc ramses}}
\def\krome    {{\sc krome}}
\def\dlsode    {{\sc dlsode}}
\def\gadget    {{\sc gadget-3}}

\begin{document}

   \title{Simulations of early structure formation:   Properties of halos that host primordial star formation}

   \author{R. Lenoble
          \inst{1}
          \and
          B. Commerçon\inst{1}
          \and
          J. Rosdahl\inst{2}
          }
 
   \institute{ENS de Lyon, CRAL UMR5574, University Claude Bernard Lyon 1, CNRS, Lyon, F-69007, France \\
              \email{romain.lenoble@ens-lyon.fr}
              \and
              University Claude Bernard Lyon 1, CRAL UMR5574, ENS de Lyon, CNRS, Villeurbanne, F-69622, France
             }

   \date{Recieved September 15, 1996; accepted March 16, 1997}

 
  \abstract
   {Population III (pop III) stars were born in halos characterised by a pristine gas composition. In such a halo, once the gas density reaches n$_{\mathrm{H}} \sim$ 1 cm$^{-3}$, molecular cooling leads to the collapse of the gas and the birth of pop III stars. Halo properties, such as the chemical abundances, mass, and angular momentum can affect the collapse of the gas, thereby leading to the pop III initial mass function (IMF) of  star formation.}
   {We want to study the properties of primordial halos and how  halos that host early star formation differ from other types of halos. The aim of this study is to obtain a representative population of halos at a given redshift hosting a cold and massive gas cloud that enables the birth of the first stars.}
   {We investigated the growth of primordial halos in a $\Lambda$CDM Universe in a large cosmological simulation. We used the hydrodynamic code \ramses{} and the chemical solver \krome{} to study halo formation with non-equilibrium thermochemistry. We then identified structures in the dark and baryonic matter fields, thereby linking the presence or absence of dense gas clouds to the mass and the physical properties of the hosting halos.}
   {In our simulations, the mass threshold for a halo for hosting a cold dense gas cloud is $\simeq 7 \times 10^5 M_{\odot}$ and the threshold in the H$_2$ mass fraction is found to be $\simeq 2 \times 10^{-4}$. This is in agreement with previous works. We find that the halo history and accretion rate play a minor role. Here, we present halos with higher HD abundances, which are shown to be colder, as the temperature in the range between $10^2 - 10^4 \, \mathrm{cm^{-3}}$ depends on the HD abundance to a large extent. The higher fraction of HD is linked to the higher spin parameter that is seen for the dense gas.
   }
   {}


   \maketitle
%

\section{Introduction}

The formation of the first stars is a critical chapter in cosmic history, marking the end of the cosmic dark ages and the dawn of the first luminous objects. Primordial stars, born from pristine gas before the onset of metal enrichment, were likely to be much more massive and short-lived than their modern counterparts, thereby influencing the subsequent evolution of the Cosmos \citep{hiranoONEHUNDREDFIRST2014}. The precise knowledge of their characteristic mass or the primordial initial mass function (IMF) is of primary importance to constrain the global evolution of the Universe and it largely remains an open question \citep{stacyBuildingPopulationIII2016}.

At early time, the Universe was nearly homogenous. Observations of the cosmic microwave background (CMB) have shown fluctuations of one part in $10^5$. These small-amplitude fluctuations grew due to gravitational instability into the first bounded objects and then into present-days galaxies. Overall, this evolution is aptly predicted by the $\Lambda$CDM model, which requires the presence of dark matter (DM). In this model, the initial primordial power-law spectrum is modified over time as a result of perturbation growth. The DM first starts to collapse and to form the first bounded objects that can accrete gas.

To collapse into stars, the baryonic gas needs to radiate away the gravitational binding energy released during the collapse. It is thus a competition between the timescales for cooling and freefall. The classic criterion from \cite{reesCoolingDynamicsFragmentation1977} states that if $t_{\mathrm{cool}} \leq t_{\mathrm{ff}}$, the gas will collapse. In pristine gas, the only available coolants are molecular hydrogen (H$_2$) and (to a lesser extent) the HD molecule. From the equality in the two timescales, a threshold in the H$_2$ fraction can be derived. With this simple theoretical framework, \cite{tegmarkHowSmallWere1997} found that population III (pop III) stars were born in the first massive bound objects of $\simeq 10^6 \, M_{\odot}$ at $z \simeq 20-30$. Today, this is still the overall framework according to recent 3D hydrodynamic simulations which follow the growth of mini-halos with masses of $10^5 - 10^6 \, \mathrm{M_{\odot}}$ in the early Universe until the formation of a collapsing cold gas clump \citep{yoshidaEarlyStructureFormation2003,hiranoPrimordialStarFormation2015}. Higher resolutions simulations using the zoom-in technique can even link the formation of halos at large scales to the birth of protostars, thereby estimating the primordial IMF \citep{hiranoONEHUNDREDFIRST2014,stacyBuildingPopulationIII2016}.

However, it has been shown that the properties of the host halo have a large impact on the enclosed star formation \citep{brommFormationFirstStars2002,hiranoONEHUNDREDFIRST2014,mckeeFormationFirstStars2008}. The angular momentum, shape, and  accretion rate of a halo (among other properties) can influence the collapse and the results from a zoom-in simulation depend a great deal on the properties of the chosen halo. Hence, it is important to study which halos lead to primordial star formation. Previous works have already pointed out this need for statistical studies. \cite{desouzaDarkMatterHalo2013} ran cosmological simulations to link the presence of a star-forming clouds with halo properties. \cite{hiranoPrimordialStarFormation2015} performed a set of cosmological simulations to derive the primordial IMF. They identified distinct populations of primordial stars, noting whether they were formed with or without a far-ultraviolet background radiation (due to the presence of nearby stars). To determine the IMF of the first stars in absence of any radiation, they used the correlation between the accretion rate at the jeans scale and the final stellar mass found in their first paper \citep{hiranoONEHUNDREDFIRST2014}.

Halo-scale angular momentum is one of the key factors determining subsequent collapse on the scale of star formation, in particular, fragmentation and disk formation. \cite{mckeeFormationFirstStars2008} studied how the accretion rate depends on the angular momentum. They concluded that the primordial IMF is set by the distribution of entropy and the specific angular momentum of the accreting gas. Similarly, simulations from \cite{brommFormationFirstStars2002} have shown that the initial angular momentum favors a disk-like structure that may even fragment. However, the relation between the angular momentum of the parent DM halo and that of the dense gas is not obvious. The baryonic gas is bounded in the gravitational potential of the parent halo. As it is falling, its angular momentum may align to that of the parent halo. \cite{hiranoPrimordialStarFormation2015} found a weak correlation between the direction of the two angular momenta, which varies with redshift. Hence, the study the of distribution of angular momentum in primordial halos is clearly valuable. It is also a key factor in galaxy formation and the halo mass function itself. \cite{druschkeShapeSpinMinihaloes2018} ran high-resolution DM-only 3D simulations of primordial halo formation to study its distribution and correlation with larger-scale properties. 

The shapes of halos can also have an impact \citep{ponAspectRatioDependence2012}. When the first stars form, halos are far from being spheroids and, instead, closer to ellipsoids, with three principal axes of different lengths. Again, \cite{druschkeShapeSpinMinihaloes2018} studied the distribution of the halo shape of halos in a DM-only 3D simulation. \cite{kazantzidisEffectGasCooling2004}, albeit their study was carried out at much lower redshifts ($z = 1-2$), showing that baryon physics have an important back-reaction on the DM distribution: taking  gas cooling into account tends to change the shape of halos dramatically, making them more spherical compared to adiabatic ones. Thus, the modelling of detailed physical processes of the gas is necessary. 

The most important process for primordial star formation is molecular formation and the path that leads to H$_2$ cooling. The large-scale physical properties of a halo can have an impact on the chemical state of the dense gas. First, there seems to be a broad consensus on the existence of a threshold in H$_2$ mass fraction at the halo scale above which a halo can host a cold collapsing gas cloud. Both theoretical works \citep{tegmarkHowSmallWere1997} and numerical simulations \citep{yoshidaEarlyStructureFormation2003} pointed at a threshold of $x_{\mathrm{H_2}} = 10^{-4} - 10^{-3}$ at the halo scale. However, it is not clear how the physical properties of the halo, in particular, the accretion rate or the angular momentum, could alter this value or trigger molecule formation. Then, at a higher density, when the central density peak reaches $10^7 \mathrm{cm^{-3}}$, \cite{hiranoPrimordialStarFormation2015} found a link between the accretion rate at the virial scale and at the Jeans length of the star-forming region. 

The role of HD cooling is debated. \cite{mcgreerImpactHDCooling2008} studied the importance of HD formation and cooling with numerical simulations. They
showed that HD cooling can be important within a halo if part of the gas is  at a low temperature and a sufficiently high density to form HD preferentially over H$_2$. They also observed that a high ionisation fraction can lead to HD formation and cooling until the CMB floor. This is also what has been observed by \cite{hiranoPrimordialStarFormation2015}: some of their halos have a much higher HD fraction, which further cools down the gas. The properties leading to a higher HD fraction are still not clear. The H$_2$ and HD abundances of the first structures depend on many factors in 3D hydrodynamic simulations, including the initial conditions of the halo, the assumed chemical network and various non-linear feedback loops \citep{klessenFirstStarsFormation2023}.

The aim of this paper is to study the properties of the first primordial star-forming halos, using a large simulation with an updated chemical network that includes HD formation. Using our simulations, we determine how halos that host collapsing gas clouds and, hence, star formation are different from other non-star-forming halos; we also consider how the properties of these star-forming halos are distributed. To do this, our simulation needs to resolve the gas, from the large-scale structure in a cosmological volume to higher density, where the gas is collapsing, until star formation occurs. We can then identify star-forming halos and get a statistically representative population of low-mass halos at high redshift. This representative catalogue of halos can be use for future studies using the zoom-in technique. \\

The outline of this paper is as follows. First, in Section \ref{section:Methodology}, we describe the numerical methodology of the hydrodynamic simulation and for the analysis and the initial conditions we use. Then, in Section \ref{section:results}, we present our results. Finally, Section \ref{section:discussion} contains a discussion, followed by  our conclusions in Section \ref{section:conclusions}.\\


\section{Methodology and numerical set-up}

\label{section:Methodology}

\subsection{Numerical methodology}

We used the adaptive mesh refinement (AMR) code \ramses{} \citep{teyssierCosmologicalHydrodynamicsAdaptive2002} to solve the interaction of DM and hydrodynamics on a 3D adaptively refined mesh. For the hydrodynamics, we used the HLLC Riemann solver \citep{toroRestorationContactSurface1994} and the MinMod slope limiter to construct gas variables at cell interfaces from their cell-centred values. To compute the gas equation of state, we used the constant adiabatic index $\gamma = 5/3$, which corresponds to an ideal monoatomic gas. This approximation is valid as the molecular fraction remains extremly low in the density range of this simulation (at most 10$^{-3}$). We used the multigrid gravity solver implemented in \ramses{} \citep{guilletSimpleMultigridScheme2011} and cloud-in-cell interpolation for DM particles.\\

The \ramses{} code uses a cubical octree structure. The cell width at the refinement level $\ell$ is $\Delta x_{\ell}$ = 0.5$^{\ell} L_{\mathrm{box}}$, where $L_{\mathrm{box}}$ is the width of the box. In our simulations, a cell is refined into eight equally sized children cells if either one of these two conditions is satisfied: \\
(a) $M_{\mathrm{DM,cell}}+\frac{\Omega_{\mathrm{m}}}{\Omega_{\mathrm{b}}}M_{\mathrm{baryon,cell}} \geq 8 \, m_{\mathrm{DM}}$, where $m_{\mathrm{DM}}$ is the mass of a DM particle, $M_{\mathrm{DM,cell}}$ and $M_{\mathrm{baryon,cell}}$ are the total DM and baryonic masses in the cell;\\
(b) the Jeans length is smaller than four cell widths, namely: if $\frac{\Delta x_{\ell}}{4} >\mathrm{\lambda_{j}}$, where $\mathrm{\lambda_{j}}=\sqrt{\frac{\pi c_{\mathrm{s}}^{2}}{G\rho}}$ with $G$ the gravitational constant, $\rho$ the gas density, $c_{\mathrm{s}}=\sqrt{\gamma k_{\mathrm{B}}T/m_{\mathrm{p}}}$ is the sound speed of the gas, with $T$ as the gas temperature, $k_{\mathrm{B}}$ the Boltzmann constant, and $m_{\mathrm{p}}$ the proton mass. \\



\subsection{Initial conditions}

We generated the cosmological initial conditions corresponding to a $\Lambda$CDM Universe with \music{} \citep{hahnMultiscaleInitialConditions2011} and the parameters as published by the Planck 2020 collaboration \citep{aghanimPlanck2018Results2020}: $\Omega_{\mathrm{m}}$ = 0.3111, $\Omega_{\Lambda}$ = 0.6889, $\Omega_{\mathrm{b}}$ = 0.04897, $\sigma_{8}$ = 0.825, H$_{0}$ = 67.66 km s$^{-1}$ Mpc$^{-1}$. Our fiducial simulation volume has a width of 1 cMpc$/h$ and 512$^3$ particles (corresponding to a minimum (coarse) level of 9). The coarse cell size is 1.9 ckpc$/h$ (150 pc at  $z$ = 18) and reaches a maximum refinement level of 19 at the last snapshot at $z$ = 18, corresponding to 0.14 pc. The DM is modelled by 512$^{3}$ particles with mass m$_{\mathrm{DM}}$ = 813 M$_{\odot}$. \\

To test the resolution convergence of this simulation, we also ran a second simulation with half the box length and the same  refinement levels and number of DM particles of 512$^3$. In this simulation, m$_{\mathrm{DM}}$ = 102 M$_{\odot}$. It also reaches a maximum refinement level of 19 at $z$ = 18. The parameters of the two simulations are shown in table \ref{table:parameter_simulation}. We discuss of the choice of the box size and the resolution in the Section \ref{subsection:halo_MF}. \\

We assumed the initial chemical abundances to be uniform in the whole volume at $z$ = 100 when we started the simulation. We computed the chemical abundances at $z=100$ with a one-zone model that takes into account the physical evolution of the Universe from $z=10^{4}$. The model details are shown in Appendix \ref{appendix:chem_initial_cdt} and the initial chemical abundances are summarised in Table \ref{table:abundance_chemical_100}.  

\begin{table*}
   \centering                          
   \begin{tabular}{ccccccc}
   \hline\hline                 
   $L_{\mathrm{box}}$ (cMpc$/h$) & $\Delta x_{\mathrm{max}}$ at $z=18$ & $\Delta x_{\mathrm{min}}$ at $z=18$ & $m_{\mathrm{DM}}\,\left(M_{\odot}\right)$ & $z_{\mathrm{ini}}$ & $z_{\mathrm{end}}$ & $M_{\mathrm{halo,\,lim}}$ \\
   \hline

   1 & 150 pc & 0.14 pc & 813 & 100 & 18 & $8.1\times10^{4}M_{\odot}$ \\

   0.5 & 75 pc & 0.07 pc & 101 & 100 & 18 & $1.1\times10^{4}M_{\odot}$\\
   
   \hline
   \end{tabular}

   \caption{Simulation volumes used in this work. The table columns are as follows, from left to right. 
   $L_{\mathrm{box}}$: volume width. 
   $\Delta x_{\mathrm{max}}$ : physical width of the coarsest grid cells.         
   $\Delta x_{\mathrm{min}}$ : physical width of the finest grid cells. 
   $m_{\mathrm{DM}}$ : mass of DM particles. 
   $z_{\mathrm{ini}}$ : initial redshift.
   $z_{\mathrm{end}}$ : redshift at the last snapshot.
   $M_{\mathrm{halo,\,lim}}$ : the resolution limit of the halo finder.
   \label{table:parameter_simulation} (see \ref{subsection:Halo_finder_clump_finder} for the proper definition)} 
\end{table*}

\begin{table}
   \caption{Chemical abundances at $z=100$}             
   \label{table:abundance_chemical_100}      
   \centering                          
   \begin{tabular}{c c}        
   \hline\hline                 
   Species  & $z=100$ \\    
   \hline                        
   E        &   2.106 $\times 10^{-4}$   \\
   H-       &   1.406 $\times 10^{-11}$  \\
   H        &   0.923                     \\
   H2       &   1.434 $\times 10^{-06}$  \\
   D        &   4.298 $\times 10^{-05}$  \\
   HD       &   2.696 $\times 10^{-10}$  \\
   HE       &   0.076                     \\
   H$^+$    &   2.106 $\times 10^{-4}$   \\
   H2$^+$   &   1.001 $\times 10^{-13}$  \\
   D$^+$    &   8.406 $\times 10^{-09}$  \\
   HEH$^+$  &   1.315 $\times 10^{-15}$  \\
   
   \hline                                   
   \end{tabular}
   \end{table}

\subsection{Chemistry and cooling}

We used the  \krome{} code \citep{grassiKROMEPackageEmbed2014} to resolve the chemistry. For each cell of \ramses{}, the non-equilibrium chemistry is computed on-the-fly in each gas cell. The system of ordinary differential equations (ODEs) describing our chemical network is solved with the \dlsode{} solver \citep{byrneStiffODESolvers1987}, which generates the sparsity structure and the Jacobian of the system.

\subsubsection{Chemical network}

We used an updated chemical network consisting of 11 species: (e$^{-}$, H, H$^+$, H$^-$, D, D$^+$, He, H$_2$, H$_2^+$, HD, and HeH$^+$) and 23 reactions. All reactions can be found in Appendix \ref{appendix:chemical_network}. The minimal network of \cite{galliChemistryEarlyUniverse1998} for H, D, and He has been updated from a comprehensive literature survey and new rate coefficients have been adopted for all 21 reactions (all reactions except H11 and H12). Simple modified-Arrhenius fits were found to adequately reproduce the literature data in the temperature range of 10-10,000 K. While differences in the fits of \cite{galliChemistryEarlyUniverse1998} do not generally exceed the investigated temperature range by more than a factor
of 2-3 , large deviations (up to a factor of 20) were observed for the formation and destruction reactions of HeH+. Full details (including the fitting coefficients) of the network will be described in \cite{FaureNetwork2023}, but they are also available upon request.

For the three-body recombination of hydrogen (3H $\rightarrow$ H$_2$ + H), the theoretical rate coefficient of \cite{forreyRATEFORMATIONHYDROGEN2013} was fitted with a modified-Arrhenius form. The Saha equation was then used to derive (and fit) the rate coefficient for the reverse reaction (A. Faure, private communication).

\subsubsection{Molecule formation}

Our chemical network is quite exhaustive, we describe here the main way in which H$_2$ is formed and destroyed. In the physical conditions of our simulations, the main formation pathway for H$_2$ is via the H$^-$ channel. In this process, electrons act as catalysts:
\begin{gather}
   \mathrm{H+e^- \rightarrow H^- +}\, h\nu  \label{chemical_reaction:Hj}  \quad \mathrm{(H1),}\\
   \mathrm{H+H^- \rightarrow H_2 + e^-} \label{chemical_reaction:H2} \quad \mathrm{(H5).}
\end{gather}
The two reactions are instantaneous as soon as one H$^-$ is formed. The other channel for H$_2$ formation in the low-density regime is the H$_2^+$ channel, but it is only dominant at very high redshifts ($z>200$), where H$^-$ is easily destroyed by CMB photons \citep[by a factor $\sim$ 100 at 1000 K]{galliDawnChemistry2013}. The three-body reaction is dominant only at really high density ($n_{\mathrm{H}} > 10^8$ cm$^{-3}$) and quickly converts the gas into fully molecular form. This reaction is included in our chemical network but we note that the simulation described in this paper never reaches densities where it becomes relevant.

At the redshift of our simulation (starting at $z = 100$), there is no efficient radiation background. The CMB photons lost most of there energy (T$_{\mathrm{rad}} \lesssim 200$ K) and there is no other source of photons prior to the formation of the first stars.

The mutual neutralisation of H$^-$ and H$^+$, which is expressed as:
\begin{equation}
   \mathrm{H^+ + H^- \rightarrow H_2} \quad \mathrm{(H6)},
\end{equation}
also forms H$_2$ but only if the ionisation fraction is non-negligible. However, in our simulation, the ionisation fraction is always low, less than $\sim 10^{-5}$, making this reaction negligible. Indeed, there is no strong ionisation mechanism at such a high redshift, such as strong shocks, high temperatures, or background radiation.\\

The reactions destroying H$_2$ are reactions (D5) and (H12). Reaction (D5) needs deuterium, which is very rare, and convert H$_2$ into HD, which is a coolant. Reaction (H12) is the collisional dissociation of H$_2$:
\begin{equation}
   \begin{aligned}
   \mathrm{D^+ + H_2 \rightarrow \,H^+ + HD} \quad \mathrm{(D5),} \\ 
   \mathrm{H_2 + H \rightarrow 3\,H} \quad \mathrm{(H12).}
   \end{aligned}
\end{equation}
It is only efficient at high density ($n_{\mathrm{H}} \geq 10^8 \mathrm{cm^{-3}}$). The number of electrons that catalysts the H$^-$ channel is thus the limiting specie for H$_2$ formation in our regime and is mainly form through reaction (H4). 

\subsubsection{Evolution of the gas temperature}
The gas temperature, $T_{\mathrm{gas}}$, depends both on the physical and chemical evolution of the gas. We followed the equation from \citet{galliDawnChemistry2013}. Expansion cooling is not included as it is already taking into account in \ramses{}. \\

The evolution of $T_{\mathrm{gas}}$  is given by:

\begin{equation}
   \begin{split}
   \frac{dT_{\mathrm{gas}}}{dt} 
   & = \frac{2}{3k_{\mathrm{B}}n}\left[\left(\Gamma-\Lambda\right)_{\mathrm{Compton}}+ \Lambda_{\mathrm{H_2}} + \Lambda_{\mathrm{HD}} + \Lambda_{\mathrm{atomic}} + \right. \\
   & \quad \qquad -{}\left.  \left(\Gamma-\Lambda\right)_{\mathrm{chem}}\right].
   \end{split}
\end{equation}

The five terms of this equation are as follows. 

\paragraph{Compton cooling:}
The first term is the Compton scattering of CMB photons on electrons \citep{galliChemistryEarlyUniverse1998}:

\begin{equation}
   \left(\Gamma-\Lambda\right)_{\mathrm{Compton}}=\frac{4k_{\mathrm{B}}\sigma_{\mathrm{T}}T_{\mathrm{rad}}^{4}\text{\ensuremath{\left(T_{\mathrm{rad}}-T_{\mathrm{gas}}\right)}}}{m_{\mathrm{e}}c}.
\end{equation}

\paragraph{H$_2$:}
We use the H$_{2}$ cooling model from \cite{gloverUncertaintiesH2HD2008}. The cooling functional form of the cooling function is

\begin{equation}
   \Lambda_{\mathrm{H_{2}}}=\frac{n_{\mathrm{H_{2}}}\Lambda_{\mathrm{H_{2},LTE}}}{1+\Lambda_{\mathrm{H_{2},LTE}}/\Lambda_{\mathrm{H_{2},n\rightarrow0}}} \, ,
\end{equation}

where $\Lambda_{\mathrm{H_{2},n\rightarrow0}}$ is the low-density limit, and $\Lambda_{\mathrm{H_{2},LTE}} = H_{\mathrm{R}} + H_{\mathrm{V}}$ the high-density limit at local thermodynamic equilibrium, sum of the vibrational and the rotational cooling: 

\begin{equation}
   \begin{aligned}
      H_{\mathrm{R}}= & \left(9.5 \times 10^{-22} T_3^{3.76}\right) /\left(1+0.12 T_3^{2.1}\right) \\
      & \times \exp \left[-\left(0.13 / T_3\right)^3\right]+3 \times 10^{-2.4} \exp \left[-0.51 / T_3\right] \\
      H_{\mathrm{V}}= & 6.7 \times 10^{-19} \exp \left(-5.86 / T_3\right) \\
      & +1.6 \times 10^{18} \exp \left(-11.7 / T_3\right) \, ,
      \end{aligned}
\end{equation}
with $T_3 = T/10^3$. 

The low-density regime is computed from the cooling $\Lambda_{\mathrm{H}_2, k}$ due to the collision with $k$ = H, H$_2$, He, H$^+$ and e$^-$ in the temperature range $10 \le T \le 6 \times 10^3$ K for H and He, and $10 \le T \le 10^4$ K for H$^+$ and e$^-$. It is expressed as:

\begin{equation}
   \Lambda_{\mathrm{H}_2, \mathrm{n} \rightarrow 0}=\sum_k \Lambda_{\mathrm{H}_2, k} \, n_k,
\end{equation}
with the coefficient from \cite{gloverUncertaintiesH2HD2008} with an ortho to para ratio of 3:1. We did not include any optically thick limit as the gas becomes optically thick above $\simeq 10^{10} \mathrm{cm^{-3}}$ and the aim of our simulation is not to reach such a density.

\paragraph{HD:} 
We use the cooling function from \cite{lipovkaCoolingFunctionHD2005} which provides a fit which depends on the temperature and density. In \krome{}, it is computed via:

\begin{equation}
   \Lambda_{\mathrm{HD}}=\left[\sum_{i, j} c_{i j} \log (T)^i \log \left(n_{\mathrm{tot}}\right)^j\right] n_{\mathrm{HD}}
,\end{equation}
with the coefficient $c_{i j}$ from \cite{lipovkaCoolingFunctionHD2005} and $n_{\mathrm{tot}}$ the total number density ($n_{\mathrm{tot}} = \sum_{i} n_i$).

\paragraph{Atomic cooling:}
Even if not included in \cite{galliDawnChemistry2013}, we include atomic cooling in our simulation. If the gas is falling without any shock, its temperature should not reach the critical temperature $\simeq 10^4$ K for atomic cooling. However, a shock could heat up the gas to temperature close to this critical value \citep{kiyunaFirstEmergenceCold2023}. We used the atomic cooling from \cite{cenHydrodynamicApproachCosmology1992}. 

\paragraph{Chemical heating and cooling:}
Chemical reactions can be exothermic or endothermic. Thus, they are either heating or cooling sources. The rate is proportional to the current rate of the reaction. For a reaction with two reactants, this is expressed as:

\begin{equation}
   \Lambda_j=E_j \, k_j \, n\left(R_{j 1}\right) \, n\left(R_{j 2}\right),
\end{equation}
where $E_j$ is the energy difference, $k_j$ is the reaction rate coefficient, while $n\left(R_{j 1}\right)$ and $n\left(R_{j 2}\right)$ are the abundance of the reactants. We took the value and the reactions listed in \cite{omukaiProtostellarCollapseVarious2000} for the parameters. The total cooling is thus a sum over all the reactions:
\begin{equation}
   \Lambda_{\mathrm{chem}}=\sum_j \Lambda_j
.\end{equation}

The H$_2$ formation is exothermic and thus a source of heat. We used the heating function from \cite{omukaiProtostellarCollapseVarious2000} which takes into account H$_2$ formation via H$^-$, H$_2^+$, and three-body reactions. The released energy is weighted by a factor, $f,$ depending on the density:

\begin{equation}
   f=\left(1+\frac{n_{\mathrm{cr}}}{n_{\mathrm{tot}}}\right)^{-1},
\end{equation}
with $n_{\mathrm{cr}}$ in $\mathrm{cm}^{-3}$ defined as:

\begin{equation}
   \begin{aligned}
n_{\mathrm{cr}}= & 10^6 T^{-1 / 2}\left\{1.6 n_{\mathrm{H}} \exp \left[-\left(\frac{400}{T}\right)^2\right]\right. \\
& \left.+1.4 n_{\mathrm{H}_2} \exp \left[-\frac{12000}{T+1200}\right]\right\}.
\end{aligned}
\end{equation}

The total chemical heating is then given as:
\begin{equation}
   \Gamma_{\text {chem }}=\Gamma_{\mathrm{H}_2, 3 \mathrm{~b}}+\Gamma_{\mathrm{H}^{-}}+\Gamma_{\mathrm{H}_2^{+}}
,\end{equation}
where the three terms correspond to the heating from three-body reactions $\left( \Gamma_{\mathrm{H}_2, 3 \mathrm{~b}} \right)$ and the H$^-$ and H$_2^+$ channel $ \left( \Gamma_{\mathrm{H}^{-}} \, \mathrm{and} \, \Gamma_{\mathrm{H}_2^{+}} \right)$, with:

\begin{equation}
   \begin{aligned}
      &\Gamma_{\mathrm{H}_2, 3 \mathrm{~b}}=4.48 f \, k_{\mathrm{H}_2, 3 \mathrm{b}} \, n_{\mathrm{H}}^3 \quad \mathrm{eV} \, \mathrm{cm}^{-3} \, \mathrm{s}^{-1},\\
      &\Gamma_{\mathrm{H}^{-}}=3.53 f \, k_{\mathrm{H}^{-}} \, n_{\mathrm{H}} \,  n_{\mathrm{H}^{-}} \quad \mathrm{eV} \, \mathrm{cm}^{-3} \, \mathrm{~s}^{-1}, \\
      &\Gamma_{\mathrm{H}_2^{+}}=1.83 f \, k_{\mathrm{H}_2^{+}} \, n_{\mathrm{H}} \, n_{\mathrm{H}_2^{+}} \quad \mathrm{eV} \, \mathrm{cm}^{-3} \, \mathrm{~s}^{-1},
   \end{aligned}
\end{equation}
where $k_j$ is the coefficient rate for the formation of the $j$ species. 

\subsection{Halo finder and clump finder}
\label{subsection:Halo_finder_clump_finder}

We saved 20 snapshots from the simulation, all of which are equally spaced logarithmically in terms of the cosmic expansion parameter from redshift $z$ = 100 to $z$ = 18. We used these outputs to identify DM and baryonic structures. We identified the DM halos by running the \adaptahop{} halofinder on each snapshot \citep{aubertOriginImplicationsDark2004,tweedBuildingMergerTrees2009}. The \adaptahop{} algorithm outputs trees of substructures in the DM field by analyzing the properties of the local density in terms of peaks and saddle points. Following the notation used in Appendix B from \cite{aubertOriginImplicationsDark2004}, we adopt N$_{\mathrm{SPH}}$ = 32, N$_{\mathrm{HOP}}$ = 16, $\rho_{\mathrm{TH}}$ = 8,  and $f_{\mathrm{Poisson}}$ = 4. Here, 
N$_{\mathrm{SPH}}$ is the number of neighbouring particles considered by the algorithm, N$_{\mathrm{HOP}}$ is the number of particles used during the smoothing step, $\rho_{\mathrm{TH}}$ is the typical overdensity used to detect halos, and $f_{\mathrm{Poisson}}$ = 4 is a statistical relevance criterion. At the end, we applied a filter that is independent of \adaptahop{} on the halo mass, rejecting all halos less massive than 100 $m_{\mathrm{DM}}$ (where $m_{\mathrm{DM}}$ is the DM particle mass) to retain only the resolved halos. We thus resolved halos down to a mass of $M_{\mathrm{vir}} \simeq 8.1 \times 10^4 M_{\odot}$ in our fiducial simulation, which is one order of magnitude below the critical mass for primordial star formation $M_{\mathrm{crit}} \simeq 4-7 \times 10^5  M_{\odot}$ identified in previous studies \citep{yoshidaEarlyStructureFormation2003,hiranoPrimordialStarFormation2015}. The critical mass of the higher resolution simulation is $M_{\mathrm{vir}} \simeq 1 \times 10^4 M_{\odot}$.

For each resolved halo, we compute the mass profile of an ellipsoid centred on the halo mass barycentre. We compute the maximum scale $\alpha_{\mathrm{crit}}$ where the virial theorem is valid with better than 20\% accuracy, namely, where this inequality is valid:
\begin{equation}
   \frac{2 E_{\mathrm{kin}, \, \alpha a,\alpha b,\alpha c} + 
   E_{\mathrm{pot}, \, \alpha a,\alpha b,\alpha c}} 
   {E_{\mathrm{kin}, \, \alpha a,\alpha b,\alpha c} + 
   E_{\mathrm{pot}, \, \alpha a,\alpha b,\alpha c}} \leq 0.2 
,\end{equation}
where $E_{\mathrm{kin}, \, \alpha a,\alpha b,\alpha c}$ and $E_{\mathrm{pot}, \, \alpha a,\alpha b,\alpha c}$ are the kinetic and potential energy contained within a spheroid of radius $\left(\alpha a,\alpha b,\alpha c\right),$ centred on the halo mass barycentre; $\left( a, b, c\right)$ are the three major axes of the halo. 
We define the virial radius as $R_{\mathrm{vir}} = \alpha_{\mathrm{crit}} \left(abc\right)^{1/3}$, based on the geometrical mean. The mass contained within this spheroid is defined as the virial mass $M_{\mathrm{vir}}$ of the halo.\\

Contrary to the DM, gas will cool and reach high density. We aim at identifying gas clumps that are likely to collapse, giving rise to pop III stars. We used the friend-of-friend (FoF) algorithm HOP algorithm \citep{eisensteinHOPNewGroupfinding1998} to identify overdensities in the gas. We extract each cell from \ramses{} to a point mass and then ran this algorithm. Following the method in the aforementioned paper, we used the recommended parameters : $N_{{\mathrm{HOP}}}$ = 16, $N_{\mathrm{merge}}$ = 4, and $N_{\mathrm{dens}}$ = 64. Here, $N_{{\mathrm{HOP}}}$ is the number of neighbours considered by the algorithm. If two clumps share more than $N_{\mathrm{merge}}$ neighbouring clumps, then they are merged; $N_{\mathrm{dens}}$ is the number of cells used to estimate the density af the central peak of each clump. We required the peak density of the clump above to be $10^4$ cm$^{-3}$ by setting the  $\delta_{\mathrm{peak}}$ parameter to this value. We  tested various values for the two other parameters $\delta_{\mathrm{out}}$ (the threshold below which a cell is not considered) and $\delta_{\mathrm{saddle}}$ (the threshold to merge clumps). This two parameters, when they are above  1 cm$^{-3}$ do not influence the detection of a clump.

We chose $\delta_{\mathrm{peak}} = 10^4$ cm$^{-3}$ as it corresponds to the density where H$_2$ cooling is saturated and the gas temperature is at its minimum $\simeq 100$ K \citep{yoshidaFormationPrimordialStars2006}. We use $\delta_{\mathrm{out}} = 10^2$ cm$^{-3}$ and $\delta_{\mathrm{saddle}} = 10^3$ cm$^{-3}$, which corresponds to the gas already cooled by molecular cooling. Once the gas reaches such a density, it will rapidly collapse and host active star formation. Previous studies use criteria based on a minimal Jeans mass and/or  a temperature threshold \citep[for example]{yoshidaEarlyStructureFormation2003}. Here, the clump detection only depends on the density threshold and the structure of the gas. We note that we tested various parameters of the clump finder, but the halos where we detect a clump are always the same. Only the multiplicity as well as the mass and/or shape of the detected clumps vary.


\section{Results}

\label{section:results}

\begin{figure}
   \centering{}
   \includegraphics[width=1\linewidth]{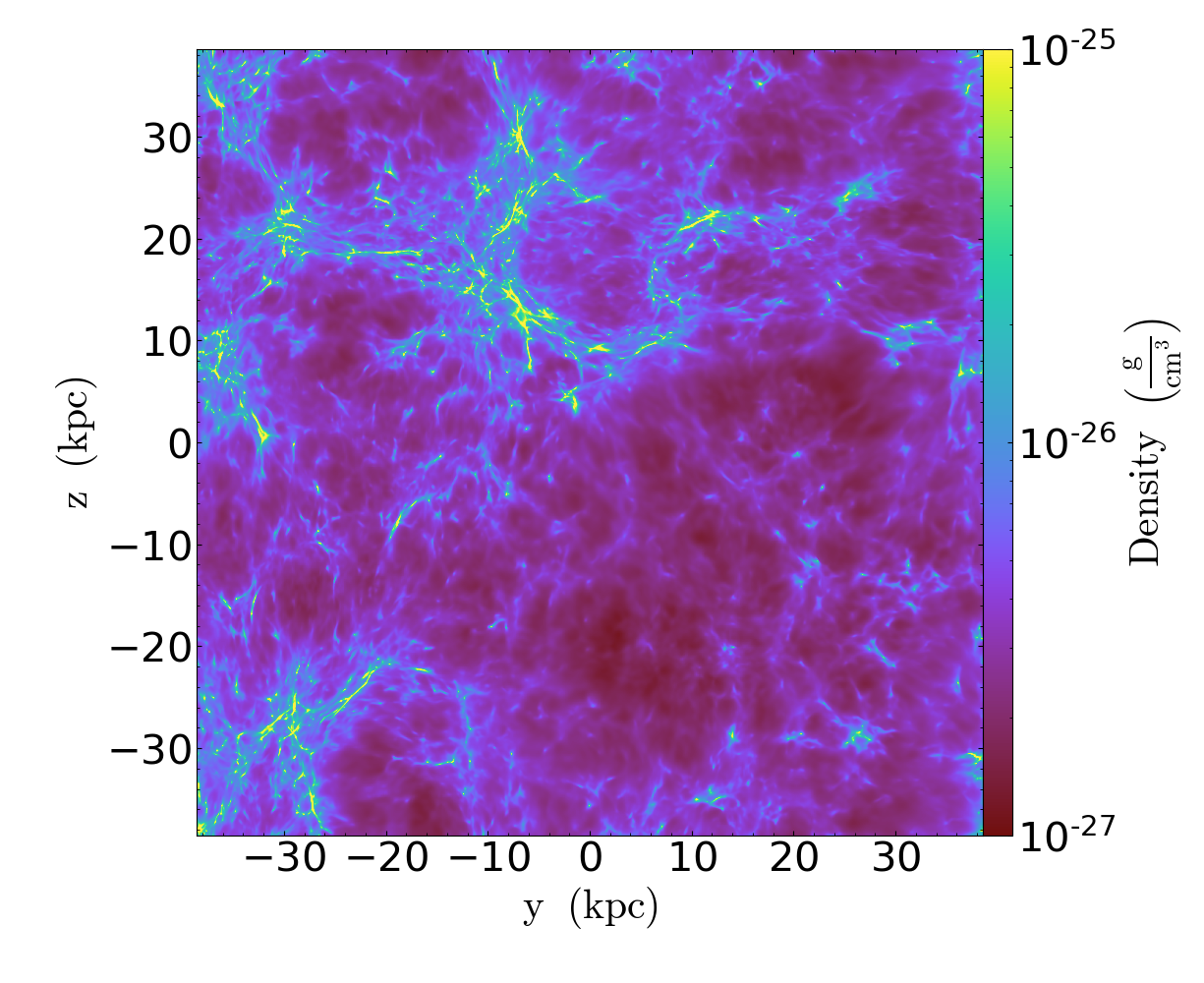}
   \caption{Mass-weighted projection of the gas density in the simulation at $z$ = 18. 
      \label{fig:all_simu}}
\end{figure}

We focus our analysis on the last snapshot of the simulation, at  $z$ = 18, as it corresponds to the peak of pop III star formation \citep{klessenFirstStarsFormation2023}. Figure \ref{fig:all_simu} shows a density map of the simulation from this snapshot. We first studied how the DM overdensities are evolving, and especially the sampling of the halo mass function in our range of interest. We then studied the links between the presence of a cold gas clump that is expected to host star formation and the physical and chemical properties of the halo. Finally, to investigate how gas is collapsing, we look at the properties of the dense gas.

\subsection{The halo mass function}
\label{subsection:halo_MF}

The halo mass function (HMF) $\mathrm{d}n/\mathrm{d}M$ is defined as the number of halos of mass $M$ per unit volume per mass interval. Depending on the primordial power spectrum and the physics of the growth of perturbation assumed, the halo mass function can be computed theoretically. The Sheth-Tormen (ST) formalism \citep{shethExcursionSetModel2002} provides an analytic function that we can compare to the mass function identified in our simulation. In particular, we first want to check that our simulation reproduces the HMF of low-mass halos ($10^5 - 10^6 \, \mathrm{M_{\odot}}$) where primordial star formation is supposed to occur \citep{yoshidaEarlyStructureFormation2003,parkPopulationIIIStar2021}. We highlight the difficulties of reproducing the halo mass function in Appendix \ref{appendix:high_res_simulation}.

\begin{figure}
   \centering{}
   \includegraphics[width=1\linewidth]{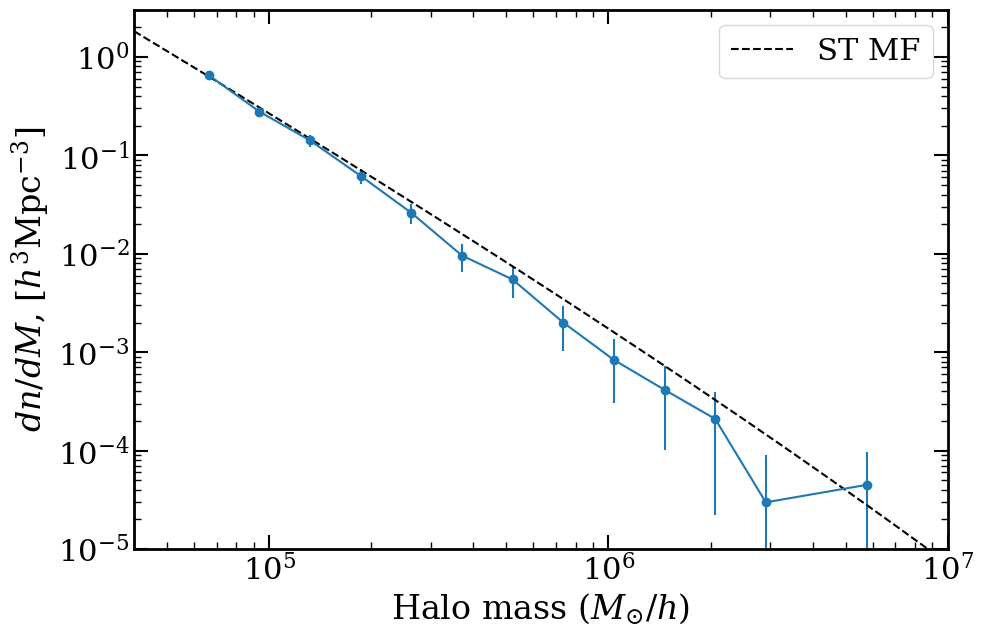}
   \caption{Comparison of the Sheth-Tormen mass function (dashed line) and the one determined in our simulation at $z = 18$ (solid line with circle). The error bars shows the Poissonian error.  The simulation is in good agreement with the analytical halo mass function.
      \label{fig:ST_VS_simulation}}
\end{figure}

First, we show a comparison between the simulated and theoretical HMF at redshift 18 in Figure \ref{fig:ST_VS_simulation}. At $z = 18$, the simulation and theory are in good agreement, particularly in the low-mass regime, down to the resolution limit ($8.1 \times 10^4$ M$_{\odot}$). There is a minor lack of halos at the high-mass end. Also, as stated in this study, the ST mass function overpredicts the abundance of rare objects at all times by up to 50\%\ compared to the simulations. Overall, the simulation reproduces  the halo mass function in the range of interest quite well, namely, between $\sim 8\times 10^4$ $M_{\odot}$ and a few $10^6$ $M_{\odot}$.

\begin{figure}
   \centering{}
   \includegraphics[width=1\linewidth]{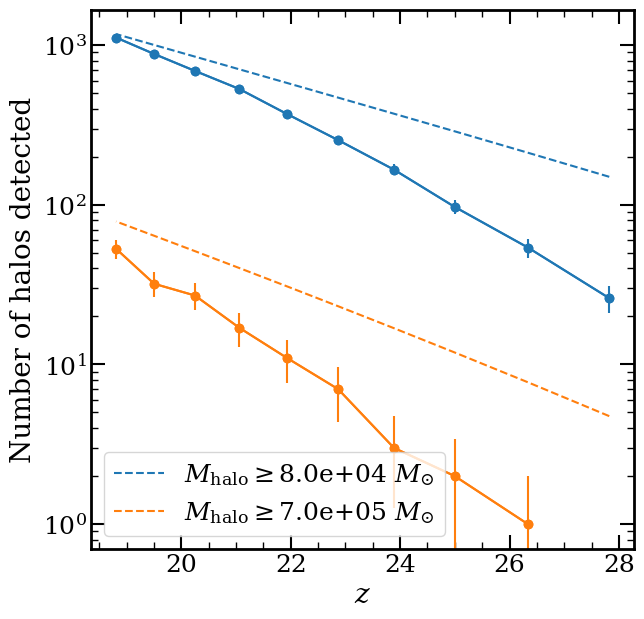}
   \caption{Number of halos detected above $8 \times 10^4 $ M$_{\odot}$ and $7 \times 10^5 $ M$_{\odot}$ in our simulation box. The dashed lines are the theoretical prediction from the Sheth-Tormen mass function, while the solid lines with circular points are the results from our simulation. The errorbars are the Poissonian error. 
      \label{fig:ST_VS_simulation_cumulative}}
\end{figure}

The second point we want to verify is whether the number of halos detected is statistically representative. Figure \ref{fig:ST_VS_simulation_cumulative} shows the number of halos detected above  $8 \times 10^4 $ M$_{\odot}$ and $7 \times 10^5 $ M$_{\odot}$. We chose these two mass thresholds as $8 \times 10^4 $ M$_{\odot}$ is our resolution limit and $7 \times 10^5 $ M$_{\odot}$ is the critical mass for a halo to host star formation found by \cite{yoshidaEarlyStructureFormation2003} and \cite{parkPopulationIIIStar2021}. The number of simulated halos converges with the theoretical expectancy with decreasing redshift. This behaviour is expected as halos of constant mass corresponds to smaller peak as redshift evolves in a box of constant size. The gap at high redshift can be explained with cosmic variance.

At $z = 18$, we detect 1376 halos more massive than $8 \times 10^4 $ M$_{\odot}$ and around 60 halos with $ M \geq 7 \times 10^5 $ M$_{\odot}$ in the fiducial simulation. Furthermore, we identified 73 clumps with the clump finder. We never detect more than one clump in a given halo. Over our sample, the mean number of cells forming a clump is 10$^5$ (the largest number is $2.6 \times 10^5$, the smallest $1.3 \times 10^4$). In the following sections, we explore in detail  why only 5\% of halos host a clump and how it mostly appears to do with the halo mass. The detection of such a high number of halos confirms that size and resolution of our our simulation are well suited for a statistical study. In Appendix \ref{appendix:high_res_simulation}, we present in a resolution study and we show that our fiducial case is convergent when  the resolution is improved.

\subsection{Clump gravitational stability}


We can now verify that by computing the virial parameter $\alpha_{\mathrm{vir}}$ whether they are gravitationally unstable or not. This parameter states whether a clump is self-supported ($\alpha_{\mathrm{vir}} \gtrsim 1$) or is collapsing  ($\alpha_{\mathrm{vir}} \lesssim 1$). For a spherical cloud with uniform density, it is expressed as:
\begin{equation}
   \alpha_{\mathrm{vir}}=\frac{5}{3}\frac{\left(c_{\mathrm{s}}^{2}+\sigma_{\mathrm{vel}}^{2}\right)\: R}{G\:M_{\mathrm{clump}}} \, ,
\end{equation}
with $c_{\mathrm{s}}$ the sound speed, $\sigma_{\mathrm{vel}}$ the velocity dispersion, $R$ the half-mass radius, and $M_{\mathrm{clump}}$ the mass of the clump (sum of the baryonic and DM mass). Sound speed and $\sigma_{\mathrm{vel}}$ are of a similar order of magnitude, between 1 and 10 km.s$^{-1}$ in the typically detected clumps. 

\begin{figure}
   \centering{}
   \includegraphics[width=0.8\linewidth]{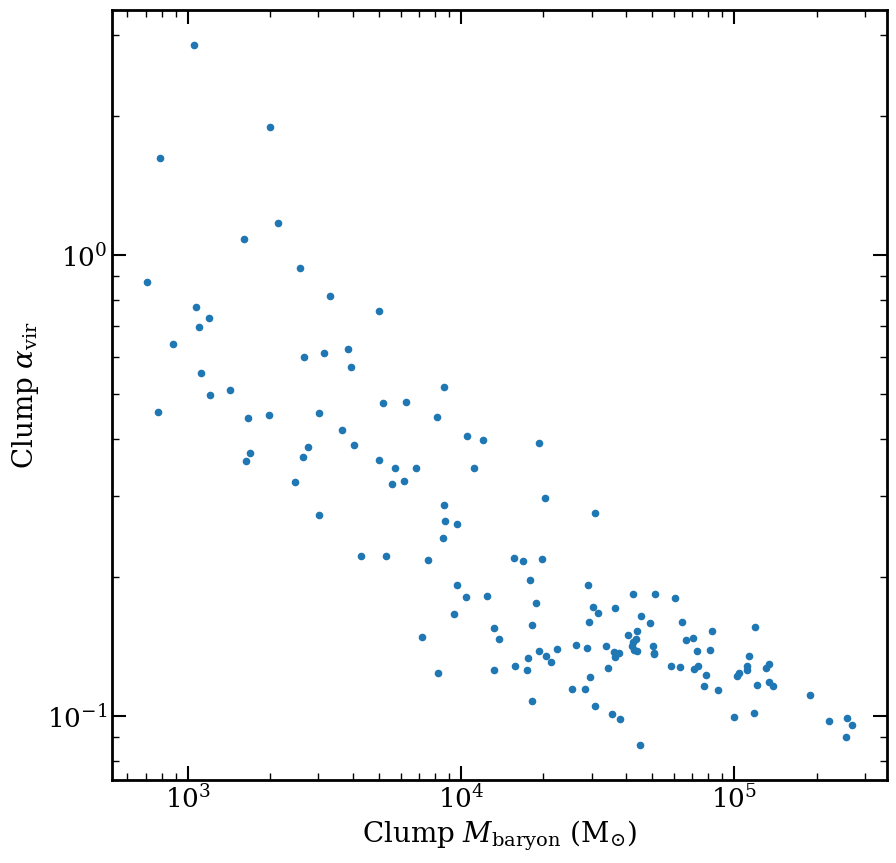}

   \caption{Clumps virial parameter $\alpha_{\mathrm{vir}}$ versus the baryonic mass at $z = 18$.
      \label{fig:m_baryon_vs_alpha}}
\end{figure}

Figure \ref{fig:m_baryon_vs_alpha} shows $\alpha_{\mathrm{vir}}$ versus the baryonic mass of each clump at $z = 18$. All the detected clumps have $\alpha_{\mathrm{vir}} \leq 3$ and the mean value is $<\alpha_{\mathrm{vir}}> \approx 0.3$. By definition, we observe a strong correlation with the mass of the clump. Indeed, the mass is the main parameter as molecular cooling sets the clump temperature. Overall, nearly all the detected clumps are gravitationally unstable, and therefore are indeed collapsing.

However, the detected clumps are not spherical (and also not with a uniform density), but using theses two assumptions are a conservative estimate. For a given cloud, $R$ and, hence, $\alpha_{\mathrm{vir}}$ would be smaller if the same mass was packed into a sphere. Also, the gas is denser at the centre of a cloud, which increases the gravitational energy compare to the uniform case. Omitting theses two assumptions would in both cases decrease $\alpha_{\mathrm{vir}}$ and would not change the conclusions of presented here.

\begin{figure*}
   \centering
   \includegraphics[width=.3\linewidth]{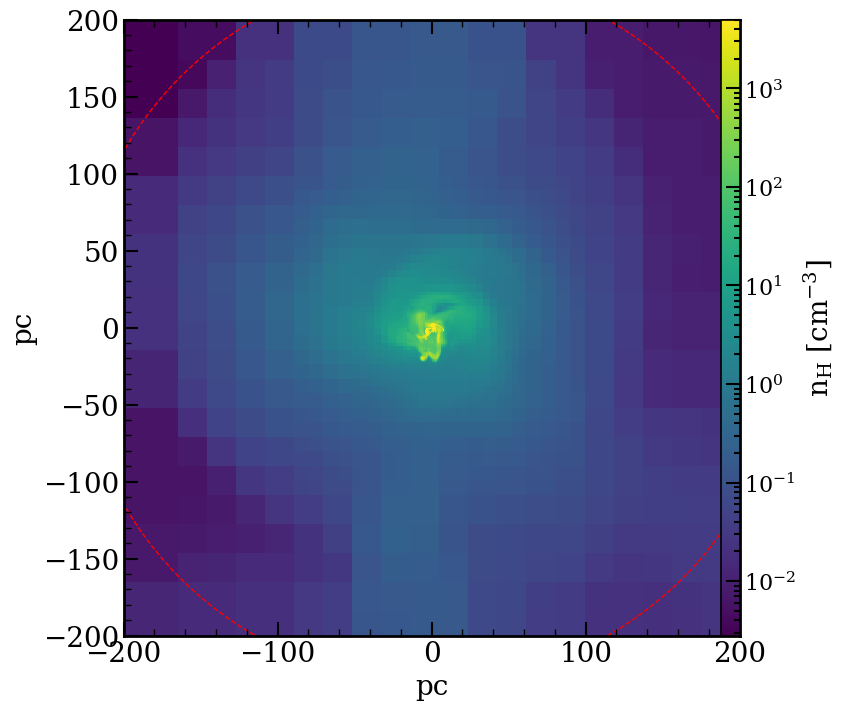}
   \includegraphics[width=.3\linewidth]{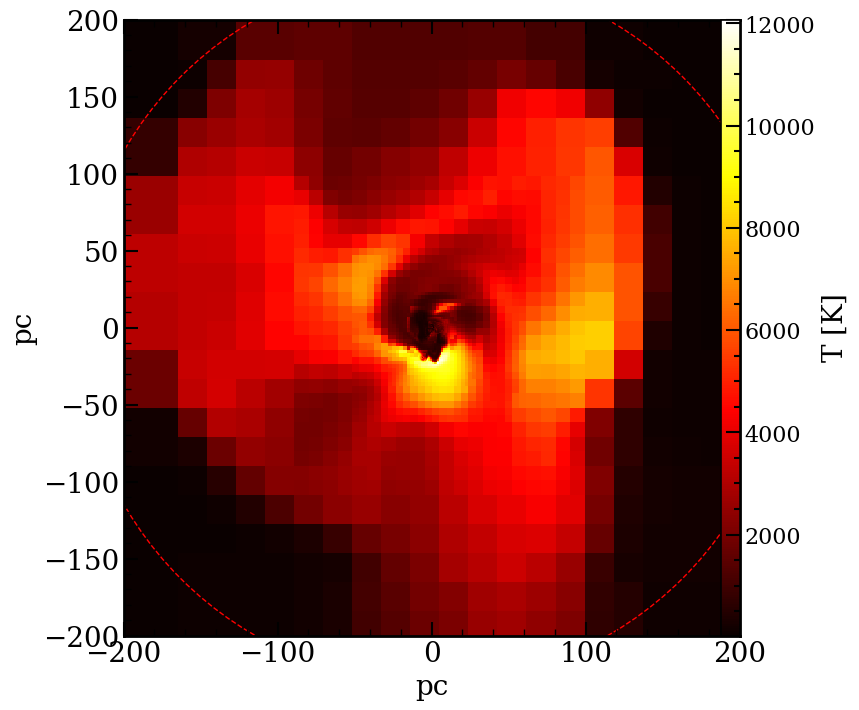}
   \includegraphics[width=.3\linewidth]{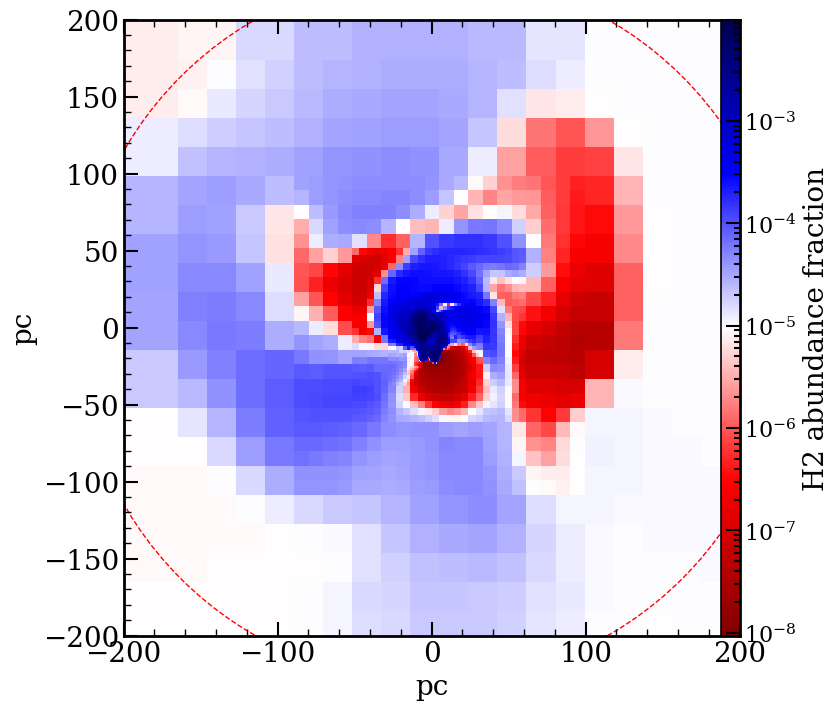}

 \caption{Maps of a halo at $z = 18$ of mass $M_{\mathrm{vir}} = 3 \times 10^6 \, \mathrm{M_{\odot}}$ and $T_{\mathrm{vir}} = 5.7 \times 10^3$ K. The first map shows the gas density, the middle one the gas temperature and the rightmost one a map of H$_2$ mass fraction. Every property is computed from a mass-average in a slice 30 pc deep. The red circle is the virial radius of the halo. The density map shows the gas collapsing from larger scale, channeling along filaments into the halo. At roughly the virial radius, the gas is heated and starts to pile up. Then, as molecule formation is triggered, gas starts to cool down. Despite the much higher density in the centre of the halo of $\sim 10^5 \mathrm{cm}^{-3}$, the gas is much colder, reaching a few hundred K. Here, in the densest region, the H$_2$ mass fraction reaches $\sim 10^{-3}$. The gas temperature correlates strongly with the H2 fraction, revealing the importance of H$_2$ for cooling.
  \label{fig:map_halo}}
\end{figure*}

\begin{figure*}
   \centering
   \includegraphics[width=.4\linewidth]{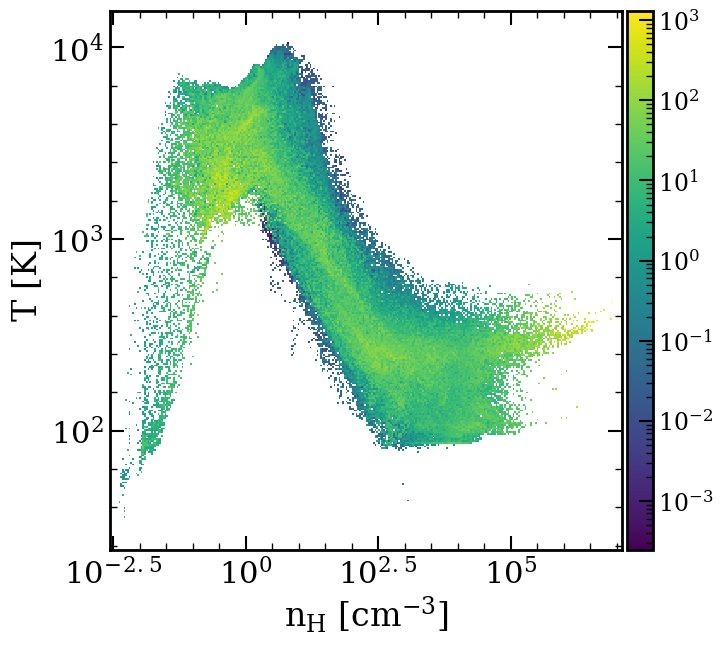}
   \includegraphics[width=.4\linewidth]{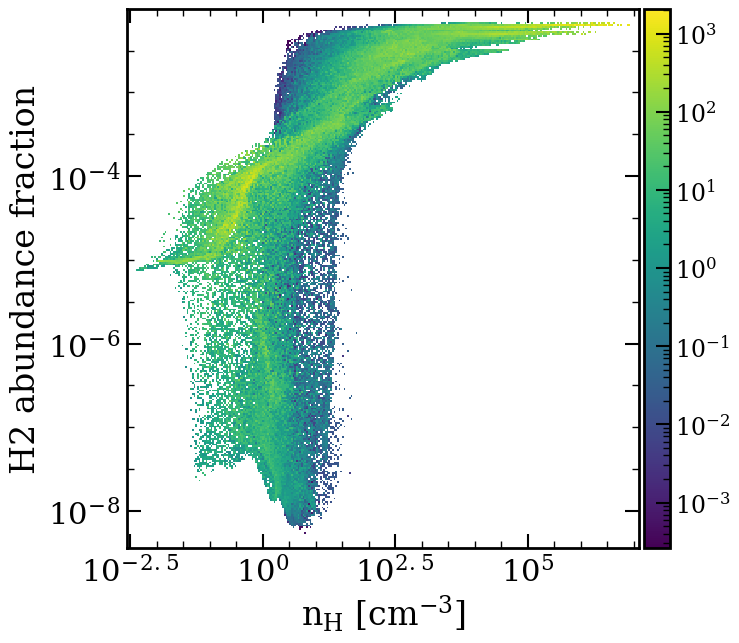} \\ 
   \includegraphics[width=.4\linewidth]{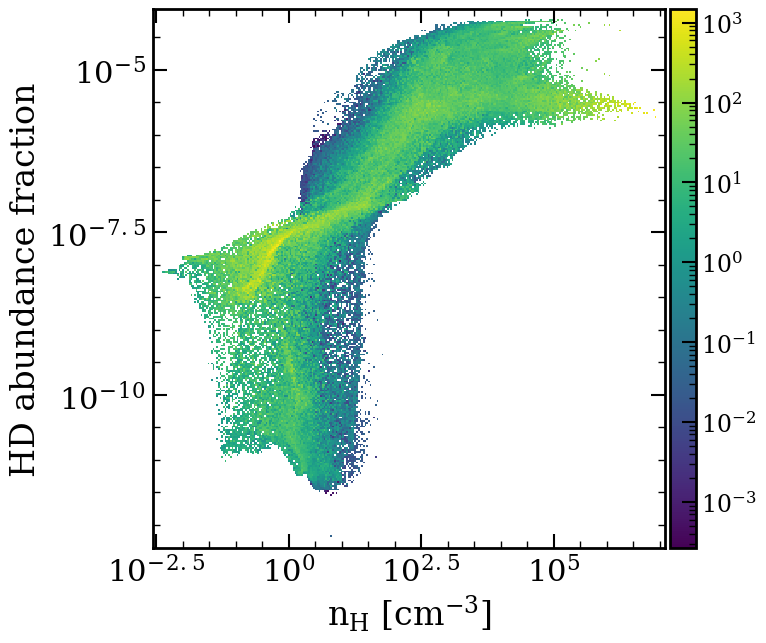}
   \includegraphics[width=.4\linewidth]{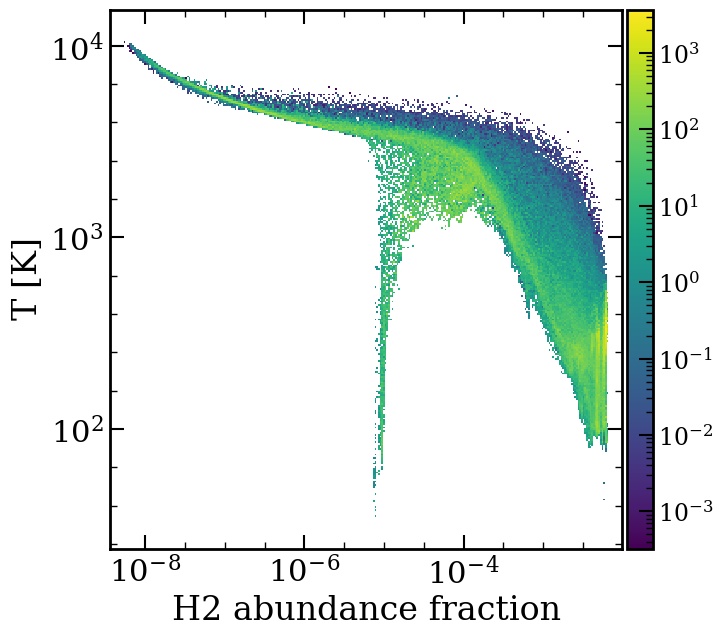}

 \caption{Phase diagrams of the gas within the virial radius of the halo of Figure \ref{fig:map_halo}.  We show the temperature, H$_2$ fraction, and HD fraction versus the gas density and the temperature versus the H$_2$ fraction. The logarighmic colormap is weighted with the mass of the gas within each bin.
  \label{fig:phase_diag}}
\end{figure*}

\subsection{Gas accretion and shocks}




As gas falls onto the halo, it is heated to its virial temperature with a mild shock. First, Figure \ref{fig:map_halo} shows maps of one collapsing halo at $z = 18$ of mass $M_{\mathrm{vir}} = 8.7 \times 10^6 \, M_{\odot}$. The gas is accreted though large-scale filaments which can be seen in the leftmost map of gas density. Then, as  collapsing molecule formation is triggered and gas starts to be enriched in H$_2$ and HD. Thus, its temperature rapidly decreases to reach a few hundred K.

Figure \ref{fig:phase_diag} shows the phase-space distribution of gas within the virial radius of one halo of mass $M_{\mathrm{vir}} = 3 \times 10^6 \, \mathrm{M_{\odot}}$ and $T_{\mathrm{vir}} = 5.7 \times 10^3$ K. The gas in the halo reaches a density of $n_{\mathrm{H}} = 10^5$ cm$^3$ and $x_{\mathrm{H_2}}$ reaches a plateau at $x_{\mathrm{H_2}} \simeq 10^3$ cm$^{-3}$ (second panel from left). The HD mass fraction is much more scattered, around $x_{\mathrm{HD}} \sim 10^{-5}$. At such densities, molecule formation enables the gas to cool down to 100 K at the most (still hotter than the CMB limit at this redshift of $T_{\mathrm{CMB}} \simeq 50$ K). The drop of $x_{\mathrm{HD}}$ at $n_{\mathrm{H}} \geq 10^{-5}$ is caused by the temperature: HD formation is more efficient at lower temperature and the temperature is still high in this halo at $n_{\mathrm{H}} = 10^5 \mathrm{cm^{-3}}$\\

The leftmost panel shows that the maximum temperature is $\approx$ 10$^4$ K, somewhat higher than the virial temperature of this halo. We find that these values are explained by a shock at smaller scales, as Figure \ref{fig:map_halo} shows. There is a shock at around 10 pc from the halo centre, in the south part of the halo. The temperature reaches more than $10^4$ K. This shock heats up the gas to this temperature as the density increases dramatically ($T \propto \rho^{\gamma -1}$). Figure \ref{fig:phase_diag} is mass-weighted, so we can see that only a small mass fraction of the gas is shocked at such high temperatures. We observe a similar behaviour in all halos more massive than $1 \times 10^6 M_{\odot}$, namely, with gas shock-heated above the virial temperature at the halo centre.

\cite{flowerThermalBalanceFirst2001} model the collapse of the initial phases of free-fall collapse in the primordial gas. They find that the infall velocity is strongly supersonic at density $n_{\mathrm{H}} = 10^2$ cm$^{-3}$. Here, we observe a  slighlty lower density  $n_{\mathrm{H}} \simeq 10^1$ cm$^{-3}$. They also observe longer H$_2$ cooling rate in the shocked region. Here, the shock is destroying H$_2$, as can be seen on the $x_{\mathrm{H_2}}$ versus $T_{\mathrm{gas}}$ panel (on the far right): the gas with the highest temperature ($T \geq 7 \times 10^3$ K) also has the lowest $x_{\mathrm{H_2}}$ ($x_{\mathrm{H_2}} \leq 10^{-6}$).

\cite{kiyunaFirstEmergenceCold2023} also observe a shock in their 3D simulation. However, in their simulation the density of the shock jumps very rapidly from  $1 - 10$ cm$^{-3}$ to $10^4 - 10^5$ cm$^{-3}$. We did not observe this behaviour in our simulation. However, we stopped our simulation at $z = 18$ and the most massive halo is $8 \times 10^6 M_{\odot}$, which is less than the critical mass for the first emergence of the cold accretion flow found in their simulation, which explains why we cannot observe the shock at high density. \\

\begin{figure}
   \centering{} 
   \includegraphics[width=1\linewidth]{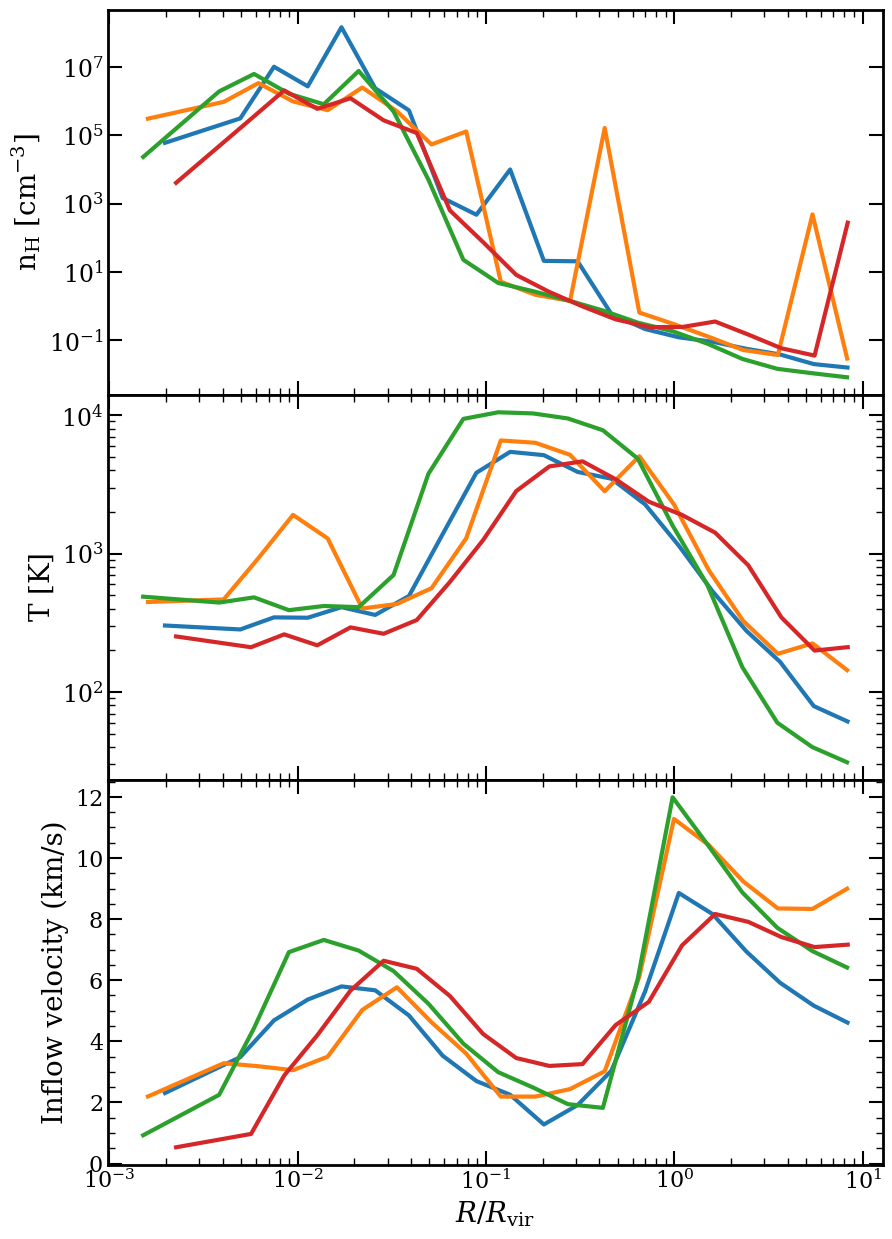}

   \caption{Profiles of mass-weighted average density, gas temperature and radial velocity of a sample of halos at $z$ = 18. The gas temperature is volume-weighted and the inflow velocity is mass-weighted. The horizontal lines are the virial properties of the halo: in the upper panel, the virial temperature; in the lower panel, the virial velocity. 
      \label{fig:profile_average}} 
\end{figure}

We also study the state of the gas as a function of the distance to the centre of the halo. Figure \ref{fig:profile_average} shows mass-weighted averaged profiles of the gas density, temperature, and inflow velocity of a few halos. The chosen halos are at $z$ = 18 and with masses of $\left[2.7,\, 3.9, \, 7.4, \, 8.8\right] \times 10^6 \, \mathrm{M_{\odot}}$ and  virial radius of $\left[2.2,\, 2.5, \, 3.1, \, 3.3 \right] \times 10^2 \, $pc, respectively. To compute the profiles, we mass-average the properties on spherical shells centred on the centre of mass of each halo. As gas is getting closer to the centre of the halo, its density increases (top panel of figure \ref{fig:profile_average}). First as there is no coolant, the temperature increases adiabatically and reaches a value close to $T_{\mathrm{vir}}$. The virial temperature of a spherical top-hat collapsing perturbation is fitted in \cite{barkanaBeginningFirstSources2001} and is written as:

\begin{multline}
   T_{\mathrm{vir}} = 1.98 \times 10^4 \left(\frac{\mu}{0.6}\right) \left[\frac{\Omega_{\mathrm{m}}}{\Omega_{\mathrm{m}}^{z}}\frac{\Delta_{\mathrm{c}}}{18\pi^{2}}\right]^{1/3} \\
      \times  \left(\frac{h \, M_{\mathrm{vir}}}{10^{8}\,\mathrm{M_{\odot}}}\right)^{2/3} \left(\frac{1+z}{10}\right)\,\mathrm{K},
      \label{eq:Tvir}
\end{multline}
with $\Omega_{\mathrm{m}}^{z}=\frac{\Omega_{\mathrm{m}}\left(1+z\right)^{3}}{\Omega_{\mathrm{m}}\left(1+z\right)^{3}+\Omega_{\Lambda}}$,  $\Delta_{\mathrm{c}}=18\pi^{2}+82d-39d^{2}$,  $d=\Omega_{\mathrm{m}}^{z}-1$, $\mu$ is the mean molecular weight (equal to 1.2 for monoatomic primordial gas), $M_{\mathrm{vir}}$ is the virial mass of the halo, $z$ is the redshift, and $h$ is the reduced Hubble constant. 
The gas then forms a bubble of hot gas with a fairly flat temperature profile. Indeed, as shown in the bottom panel, the radial velocity drops sharply. This hot phase of the gas can be seen in Figure \ref{fig:map_halo} and is deficient in H$_2$ which impeaches the gas from cooling. Then, at a higher density, molecule formation starts to enable the gas to cool down but the size of the dense gas cloud is too small (the half-mass radius is between 5 and 15 pc), seen in Figure \ref{fig:profile_average}.

In the velocity profiles (bottom panel of Figure  \ref{fig:profile_average}), we compare against the expression from  \cite{barkanaBeginningFirstSources2001}. They compute the velocity from the virial theorem as:

\begin{equation}
   V_{\mathrm{vir}} = 23.4  \left[\frac{\Omega_{\mathrm{m}}}{\Omega_{\mathrm{m}}^{z}}\frac{\Delta_{c}}{18\pi^{2}}\right]^{1/6} \\
   \times  \left(\frac{h \, M_{\mathrm{h}}}{10^{8}\,\mathrm{M_{\odot}}}\right)^{1/3} \left(\frac{1+z}{10}\right)^{1/2}\,\mathrm{km \, s^{-1}}
   \label{eq:Vvir}
,\end{equation}
with the same parameters as in Equation (\ref{eq:Tvir}).
At large scales, the gas is falling toward the halo and its velocity nearly reaches the virial velocity. Then, just inside the virial radius, the radial velocity drops sharply.

\subsection{Properties of star-forming primordial halos}
\label{subsection:properties_halos}

We now investigate the properties of the star-forming halos in our simulation with the aim of finding statistically relevant criteria for halos hosting clumps, both in term of physics and chemistry. Previous works highlight a critical threshold in molecular abundance and a mass threshold. We seek to confirm theses results with our simulation. 

\subsubsection{Chemical properties of halos hosting a cold gas clump}

The only chemical species likely to cool the gas are H$_2$ and HD. As the abundance of HD is negligible compared to H$_2$,    cus on the molecular hydrogen number fraction $x_{\mathrm{H_2}}$, which is expected to determine the onset of cooling. We compute $x_{\mathrm{H_2}}$ with a mass weighted-average of the gas within the virial radius. Before the onset of molecular cooling, the temperature of the gas is set by the virial temperature of the halo which can be computed analytically. The virial temperature is computed from Equation (\ref{eq:Tvir}).

\begin{figure}
   \centering{} 
   \includegraphics[width=0.8\linewidth]{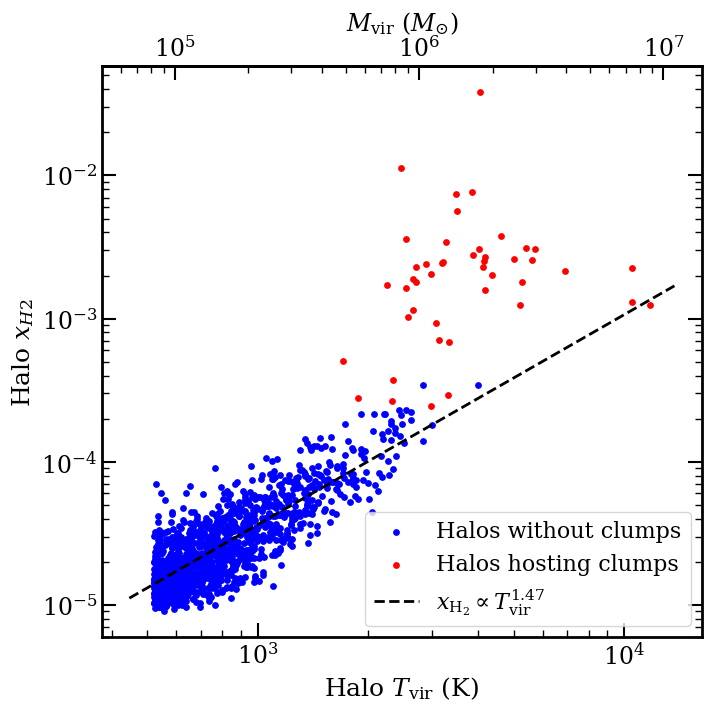}

   \caption{Mass-weighted H$_2$ fraction, $x_{\mathrm{H_2}}$, versus virial the temperature for all halos. The red circles are halos hosting clumps and the blue ones are those without them. The black dashed line is a power law fit for the blue points, i.e. halo without clumps. 
      \label{fig:H2_fract_vs_Tvir}}
\end{figure}

We plot the mean H$_2$ mass fraction versus the virial temperature in Figure \ref{fig:H2_fract_vs_Tvir}. Red circles correspond to halos hosting cold gas clumps detected by the HOP algorithm and the blue to the ones without a clump. We clearly identify a threshold $x_{\mathrm{H_2, crit}} \simeq 2 \times 10^{-4}$ above which a halo is always hosting a clump, in agreement with \citep{yoshidaEarlyStructureFormation2003, tegmarkHowSmallWere1997}. 

Figure \ref{fig:H2_fract_vs_Tvir} shows two different trends. First $x_{\mathrm{H_2}}$ increases quite regularly with $T_{\mathrm{vir}}$, as the halo is accreting mass until $\simeq 2-3 \times 10^3$ K, which corresponds, at this redshift, to virial masses of $\simeq 8 \times 10^5 - 10^6 M_{\mathrm{vir}}$. We fit with a power law $x_{\mathrm{H_2}}$ versus $T_{\mathrm{vir}}$ only for the halos that do not host clumps, namely, the blue points. We find that $x_{\mathrm{H_2}}$ scales as $T^{1.47}$, in fairly good agreement with the scaling of $T^{1.52}$ found by \cite{yoshidaEarlyStructureFormation2003} and \cite{tegmarkHowSmallWere1997}. The computed slope depends on the chemical network used and can be computed analytically. \cite{tegmarkHowSmallWere1997} model H$_2$ formation via the H$^-$ channel as a function of the temperature of the gas which is equal to the halo virial temperature before the onset of molecular cooling. The H$^-$ channel is the main route of H$_2$ formation in our physical conditions. Its effective formation rate, $k_{\mathrm{H_2}}$, depends on the coefficient rate of the reaction forming H$_2$ divided by the one destroying H$^-$. Following \cite{tegmarkHowSmallWere1997} in assuming that the molecular fraction is $x_{\mathrm{H_2}} \ll 1$, it depends only on the gas temperature and is expressed as:
\begin{equation}
   x_{\mathrm{H_2}} \simeq \frac{k_{\mathrm{H_2}}}{k_{\mathrm{H1}}(T_{\mathrm{gas}})} = \frac{1}{k_{\mathrm{H1}}(T_{\mathrm{gas}})} \times \frac{k_{\mathrm{H4}}(T_{\mathrm{gas}}) k_{\mathrm{H5}}(T_{\mathrm{gas}})}{(k_{\mathrm{H5}}(T_{\mathrm{gas}}) + k_{\mathrm{H3}}(T_{\mathrm{rad}})/n_{\mathrm{H}})}, 
\end{equation} 
where $k_i$ in the rate of reaction $i$ which depends either on the gas temperature or on the radiation temperature for the photoionisation reactions. In the range of temperature between 10 K and $10^4$ K, this rate can really close to be proportional to $T^{1.52}$, in agreement with the observed rate in our simulation. 

Then, when a halo starts to host a clump, the scatter in the plot increases dramatically and the relation with $T_{\mathrm{vir}}$ breaks. Indeed, the assumption that the molecular fraction is negligible is no longer valid. Also, the gas temperature is no more set by the virial temperature as H$_2$ cooling starts to play a role. This trend was not observed in \cite{yoshidaEarlyStructureFormation2003} as they find that $x_{\mathrm{H_2}}$ continues to increase regularly, independently of the presence or absence of a clump. Recently, \cite{reganMassiveStarFormation2022} who performed a numerical simulation of the first halos also observe a similar trend as here: the scatter in the cooling time increases above a few times $10^6 \, \mathrm{M_{\odot}}$, similar to our simulation. 

\begin{figure}
   \centering{}
   \includegraphics[width=1\linewidth]{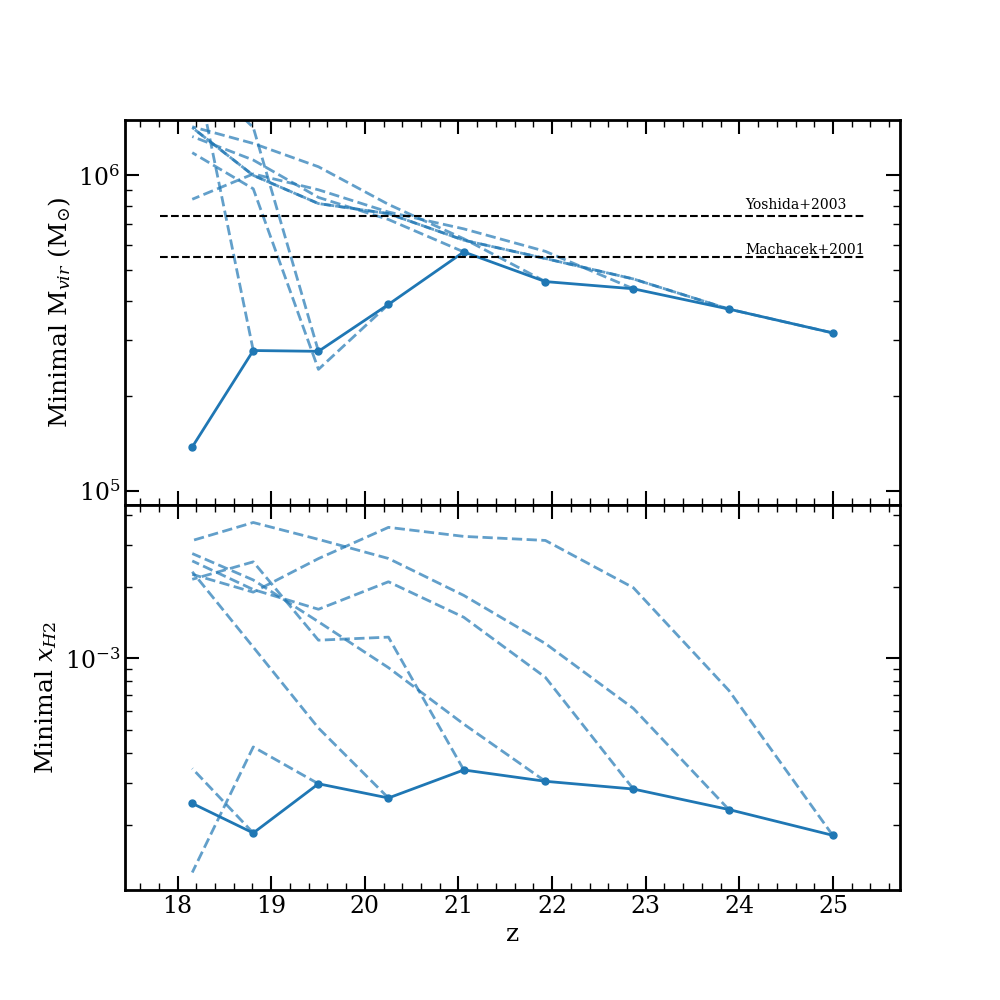}
   \caption{Minimum mass (top) and minimum $x_{\mathrm{H_2}}$ fraction (bottom) of halos hosting clumps in our simulation as a function of redshift. The blue solid lines show the evolution of these minimums and the dashed lines show the subsequent mass or $x_{\mathrm{H_2}}$ evolution of the relevant 'minimal' halo, usually increasing in both mass and H2 fraction. In the top panel, the horizontal lines show the critical mass found by \cite{yoshidaEarlyStructureFormation2003} and \cite{machacekEffectsSoftXray2003}.
   \label{fig:Min_H2_mass_halo}}
\end{figure}

Figure \ref{fig:H2_fract_vs_Tvir} only reports one simulation snapshot, at $z = 18$. We investigate whether the critical $x_{\mathrm{H_2}}$ identified is independent of redshift. The lower panel of Figure \ref{fig:Min_H2_mass_halo} shows the evolution over redshift of the minimum H$_2$ mass fraction of halos hosting clumps. The minimal $x_{\mathrm{H_2}}$ does not depend on the redshift and is roughly constant at $\simeq 2-3 \times 10^{-4}$. The dashed lines show the evolution of individual halos, as we trace them forward across snapshots. For these individual halos, the H$_2$ fraction increases and reaches an upper limit of $\simeq 2-3 \times 10^{-3}$. This upper limit is probably due to $x_{\mathrm{H_2}}$ being computed at the virial scale and to the fact that the simulation does not resolve the densest gas, which is expected to become fully molecular above $10^{10} \mathrm{cm^3}$. 

\subsubsection{Physical properties of halos hosting cold gas clump}
\label{subsection:halo_phys_prop}

The minimal mass, $M_{\mathrm{crit}}$, of a halo hosting star formation is the main property we investigate here. This critical mass has been inferred from theoretical predictions \citep{tegmarkHowSmallWere1997}. It can be used to estimate the global primordial star formation rate. For example, \cite{chantavatMostMassivePopulation2023a} use a redshift-dependant equation to compute the abundances of Pop III. In their 3D simulation, \cite{yoshidaEarlyStructureFormation2003} investigate the minimal mass and find $M_{\mathrm{crit}} = 7 \times 10^5 $ M$_{\odot}$, independent from redshift. Similarly, \cite{machacekSimulationsPregalacticStructure2001c} find a critical mass of $\simeq 3 \times 10^5 \, \mathrm{M_{\odot}}$ in a 3D simulation.

The upper panel of Figure \ref{fig:Min_H2_mass_halo} shows the minimum mass of halos hosting gas clumps in each output. We infer a minimum mass of $\sim 1-5 \times 10^5$ M$_{\odot}$, which only weakly depends on the redshift. Our higher resolution simulation agrees well with this result from the fiducial simulation, as we show in Appendix \ref{appendix:high_res_simulation}. The value we find is in fairly good agreement with previous simulations, as the value found here is between the results from \cite{yoshidaEarlyStructureFormation2003} and \cite{machacekSimulationsPregalacticStructure2001c}. 

\begin{figure}
   \centering{}
   \includegraphics[width=0.9\linewidth]{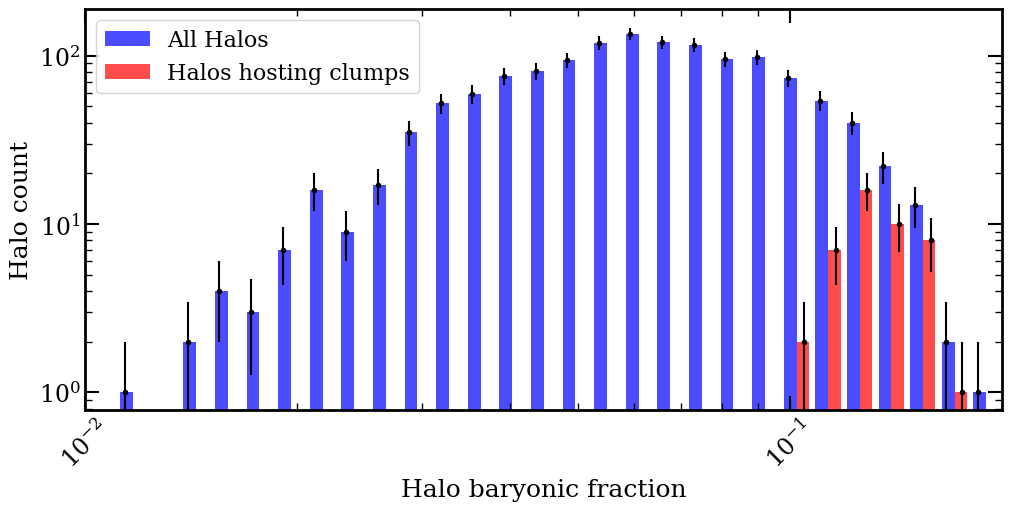}
   \caption{Histogram of the probability density of the baryonic fraction for the all population of halos (blue bars) and those hosting clumps (red bars). Error bars correspond to the Poissonian error.
   \label{fig:hist_fb}}
\end{figure}

The baryonic fraction, $f_{\mathrm{b}}$, is also expected to have an influence. Figure \ref{fig:hist_fb} shows a histogram of the baryonic fraction for the all sample of halos (blue) and those with clump (red) with the Poissonian error bars. We observe a clear threshold of $f_{\mathrm{b, crit}} \sim 10^{-1}$ for a halo to host a clump.

\subsection{Halo formation history and accretion rate}


The presence of a clump is first dictated by the virial mass (or, alternatively, the molecular fraction $x_{\mathrm{H_2}}$, which is also correlated to virial mass, see Figure \ref{fig:H2_fract_vs_Tvir}). However, there is a range of masses where there exist halos with and without clumps, therefore, we want to look at other halo properties that can distinguish between these two classes of halos in the mass range where there is an overlap. 

We first focus on the halo history and accretion rate. Indeed, the growth of halos via accretion and mergers can be violent and thus have an impact on their properties and star formation. If slow accretion tends to favor equilibrium within the halos, a merger of two DM halos can affect the thermal and chemical evolution of the gas. To study the effects of accretion and mergers, we follow the evolution of individual halos and compare those that form clumps and those that do not.

\subsubsection{Halo history}

\begin{figure}
   \centering{} 
   \includegraphics[width=1\linewidth]{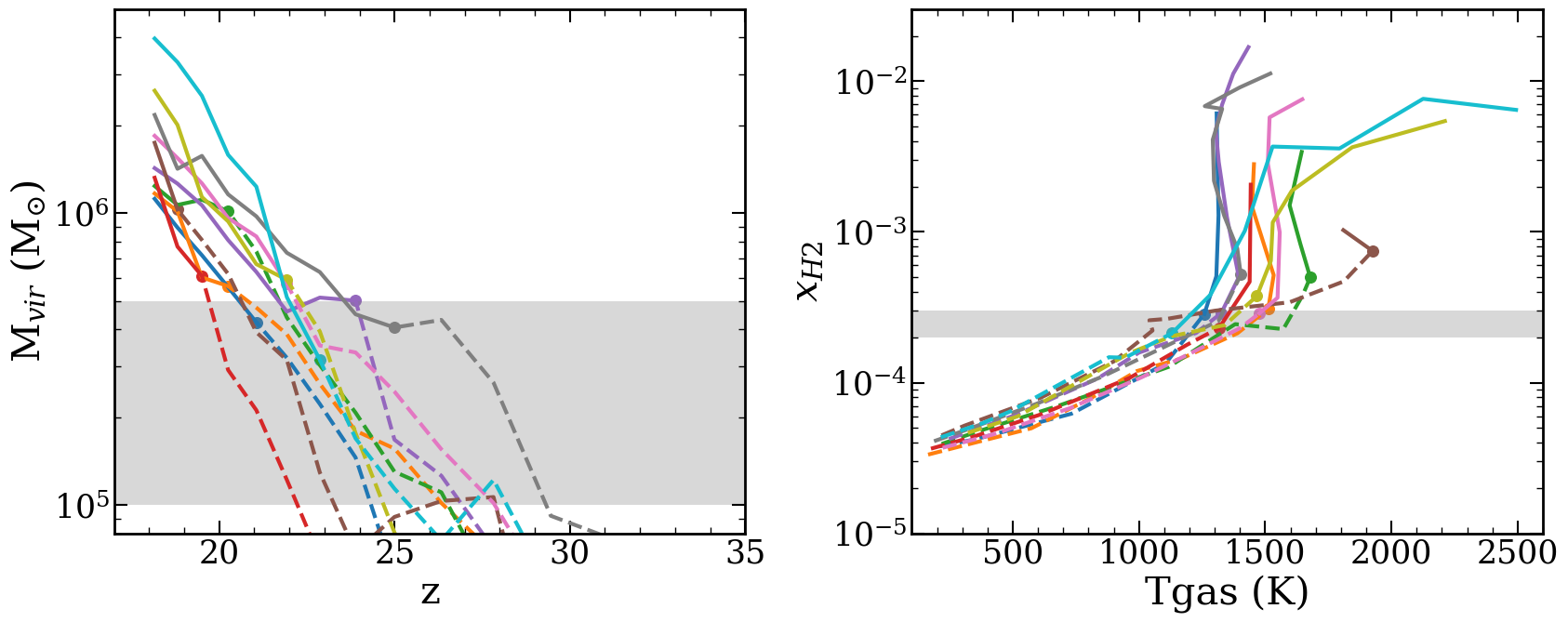}
   \includegraphics[width=1\linewidth]{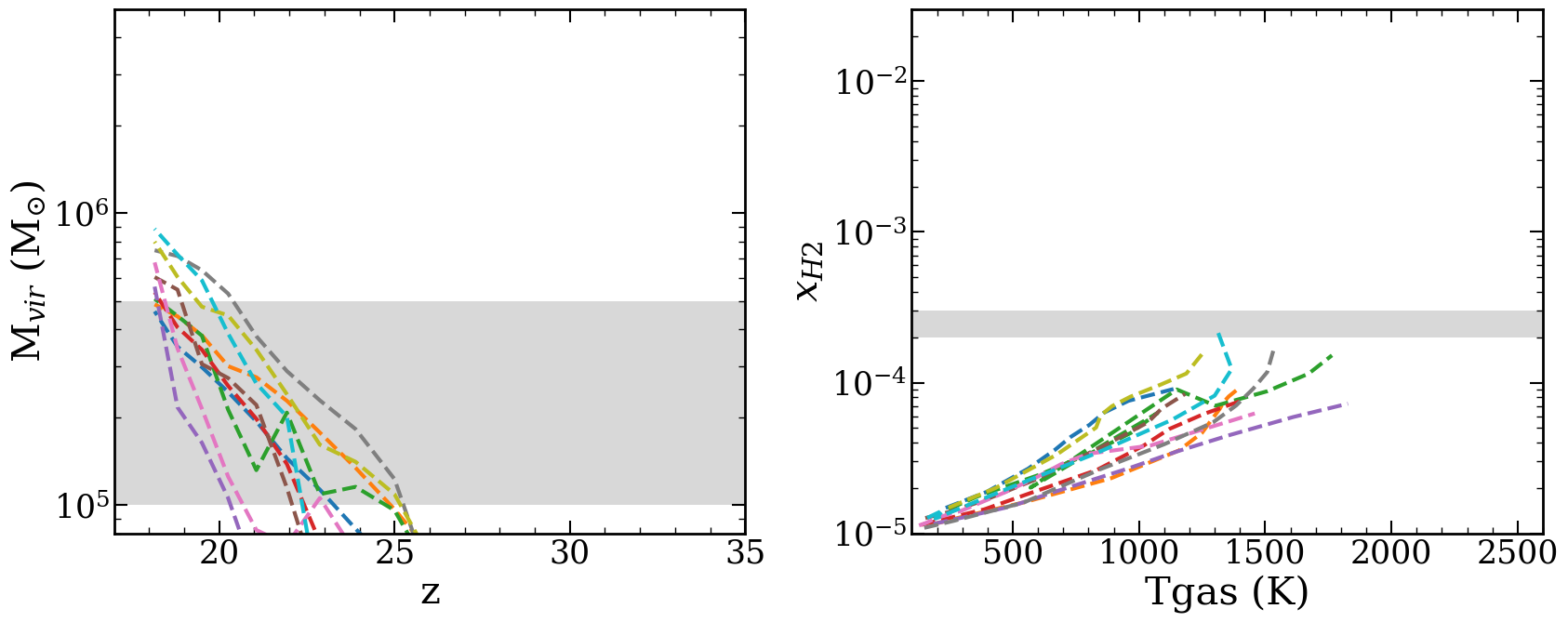}

   \caption{Evolution of halos across snapshots. Top: Sample of halos hosting clump at $z = 18.8$, Left : Halo virial mass vs redshift, Right: Halo H$_2$ abundance vs mass average gas temperature. The boundary between the solid line and dashed line corresponds to the snapshot at which a halo first host a cold gas clump. Bottom: Same plot for a sample of halos that do not host cold gas clump. The gray rectangle shows the interval in the crirical value of $M_{\mathrm{vir}}$ and $x_{H_2}$ identified in Section \ref{subsection:properties_halos}.
      \label{fig:Halo_evol}}
\end{figure}

We can study the evolution of halos to see if a higher accretion rate, which heats the gas, delays the presence of a clump. Figure \ref{fig:Halo_evol} shows the evolution of a random sample of halos hosting clumps at $z = 18$ (top row) and another subset of halos that do not (bottom row). For both, the right panel shows the evolution of the virial mass as a function of the redshift and the left one gives the H$_2$ abundance as a function of the volume average gas temperature. In the top row, the boundary between the solid and the dashed circles marks the first snapshot at which a halo hosts a clump. The first row reflects Figure \ref{fig:Min_H2_mass_halo} as we can clearly see a threshold in both the virial mass and $x_{\mathrm{H_2}}$ above which a halo hosts a clump. The threshold is much clearer for $x_{\mathrm{H_2}}$ as there is a wider range in $M_{\mathrm{vir}}$ for which there exists halos both with and without gas clumps. None of the halos without cold gas clumps reaches either of these thresholds ($M \geq 7 \times 10^5 \mathrm{M_{\odot}}$ or $x_{\mathrm{H_2}} \geq 2 \times 10^{-4}$).

The plot of $x_{\mathrm{H_2}}$ versus the gas temperature highlights three phases in the halo evolution. First, as the halo accretes mass and is collapsing, its temperature increases adiabatically as it cannot exchange heat with the surrounding medium. Then, once it reaches the critical temperature for H$_2$ formation ($\sim 1-1.5 \times 10^3$ K), H$_2$ starts to play a role and to cool down the gas. However, at the same time, the halo is still accreting gas which acts as a heating term, counterbalancing the effect of H$_2$ cooling. This explains the vertical part of the trajectory. Finally, H$_2$ formation starts to saturate and the halo heats up again as cooling cannot compensate dynamical heating. Overall, the halo history seems to  have only a limited impact.



\subsubsection{Accretion rate} 

\begin{figure}
   \centering{} 
   \includegraphics[width=0.8\linewidth]{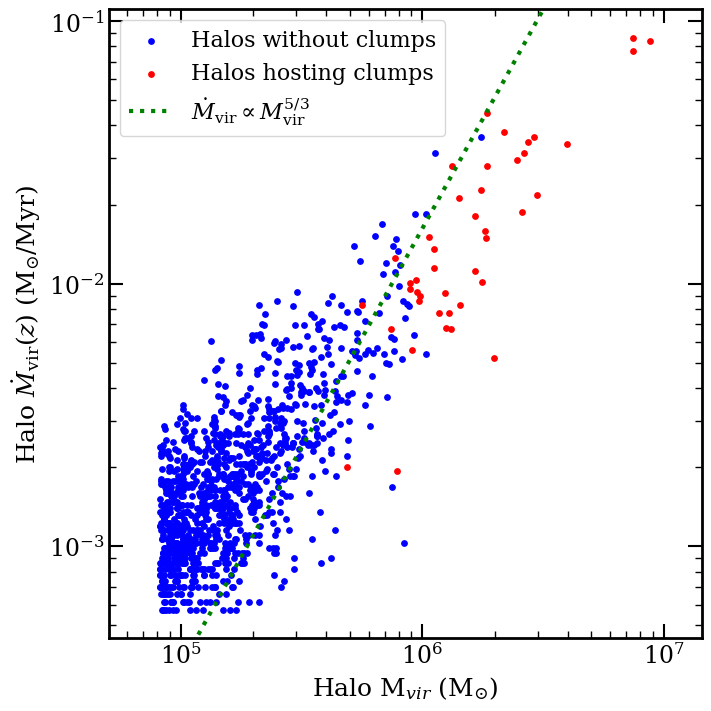}

   \caption{Virial mass versus the accretion rate at the virial radius at $z$ = 18. The red circles are halos hosting clumps and the blue ones those without. The green dotted line shows $\dot{M}_{\mathrm{crit}} \propto M_{\mathrm{vir}}^{5/3}$. 
      \label{fig:acc_rate_mvir}} 
\end{figure}

We can also directly study the impact of the accretion rate in a given snapshot. For this, we show in Figure \ref{fig:acc_rate_mvir} the accretion rate at the virial radius versus the virial mass of halos. We computed the accretion rate with the difference of mass of each halo between two snapshots. Again, the red points correspond to halos that host cold and dense clumps and the blue ones those that do not. The accretion rate is computed between two snapshots at $z$ = 18 and $z$ = 18.8. First, we can clearly see that at a given mass, halos that host cold gas clumps tend to have a smaller accretion rate. This is especially true at the critical mass close to $\simeq 10^6$ M$_{\odot}$ where a halo first hosts a clump. 

Following \cite{yoshidaEarlyStructureFormation2003}, we computed a critical accretion rate by comparing the heating due to the accretion to the H$_2$ cooling. The dynamical heating due to the accretion rate can be linked to the increase of the virial temperature of the halo as follows:

\begin{equation}
   \frac{dQ_{\mathrm{dyn.\,heating}}}{dt} = \frac{k_{\mathrm{B}}}{\gamma-1} \frac{dT}{dt} = \alpha M_{\mathrm{vir}}^{-1/3}\frac{dM_{\mathrm{vir}}}{dt}, 
\end{equation}
where we use Equation (\ref{eq:Tvir}) for the second equality, $\alpha$ depends on the factors of this equation. The temperature of the gas could be different from the virial temperature, but the two are equal before the onset of H$_2$ cooling. 
The H$_2$ cooling rate is:
\begin{equation}
   \frac{dQ_{\mathrm{H_{2}}}}{dt} = \Lambda_{\mathrm{H_{2}}}\times f_{\mathrm{H_{2}}},
\end{equation}
where $\Lambda_{\mathrm{H_2}}$ is the molecular hydrogen cooling rate. \\
If the cooling and heating terms are equal, a halo cannot cool down and, thus, clump formation is delayed. The molecular cooling rate only depends on the abundance of H$_2$, which can be assumed constant in our restricted mass range and $\Lambda_{\mathrm{H_{2}}}$, which is roughly proportional to $T^2$ between 100 and 5000 K as we use the cooling rate from \cite{gloverUncertaintiesH2HD2008}. Equalling the two rates gives a critical accretion rate above which a cloud cannot cool down thanks to molecule cooling. As $T_{\mathrm{vir}} \propto M_{\mathrm{vir}}^{2/3}$, the crirical growth rate reads as $\dot{M}_{\mathrm{crit}}=\frac{1}{\alpha}M_{\mathrm{vir}}^{1/3}\Lambda_{\mathrm{H_{2}}}f_{\mathrm{H_{2}}}\propto M_{\mathrm{vir}}^{5/3}$. Figure \ref{fig:acc_rate_mvir} shows that this relation is valid above $5 \times 10^5$ M$_{\odot}$ but breaks for less massive halos. Indeed, in this case, the limiting factor is the mass of the halo.

\subsection{Spin parameter}


We computed the angular momentum $\overrightarrow{J}$ with the classic expression:
\begin{equation}
   \label{eq:AM}
   \overrightarrow{J}=\underset{r_{i}}{\sum}m_{i}\overrightarrow{r_{i}}\times\overrightarrow{v_{i}},
\end{equation}
with $\overrightarrow{r_{i}}$ as the position of the DM particle or centre of gas cell, $i,$ with respect to the clump barycentre (which is not necessary the same as the halo barycentre), $m_{i}$ is the mass of the particle or cell, and $\overrightarrow{v_{i}}$ its velocity relative to that    of the clump. We compute $\overrightarrow{J}$ both for the DM and the baryons.

The angular momentum depends sensitively on the mass of an object, as Equation (\ref{eq:AM}) is a sum. Moreover, more massive objects have higher rotational velocities. Therefore, to compare the rotational support of halos with a wide range of mass, we would need a more appropriately weighted physical quantit : we chose to compute the spin parameter, following the definition from \cite{bullockUniversalAngularMomentum2001}, 

\begin{equation}
   \lambda=\frac{J_{\star}}{\sqrt{2} R M_{\star} V_{\mathrm{circ}}} \, ,
\end{equation}
with $V_{\mathrm{circ}}=\sqrt{\frac{G M}{R}}$, where $G$ is the gravitational constant, $M$ is the mass of the object, and $R$ is the radius in case of a circular object. Also, $J_{\star}$ and $M_{\star}$ are the angular momentum in the halo frame and the mass of the considered particles (either the gas or the DM in our case), $J_{\star}/M_{\star}$ is the specific angular momentum. 
For the DM, we computed the spin parameter using the virial properties $R_{\mathrm{vir}}$ and $M_{\mathrm{vir}}$. For the baryons, we computed the angular momentum and the mass over all cells within a sphere centred of the centre of mass and of radius, $R_{\mathrm{vir}}$. Numerical studies \citep{brommFormationFirstStars2002,hiranoONEHUNDREDFIRST2014} have highlighted the importance of the angular momentum and, thus, of the spin parameter, as both of these aspects may influence the collapse of the gas cloud.

\begin{figure}
   \centering{}
   \includegraphics[width=1\linewidth]{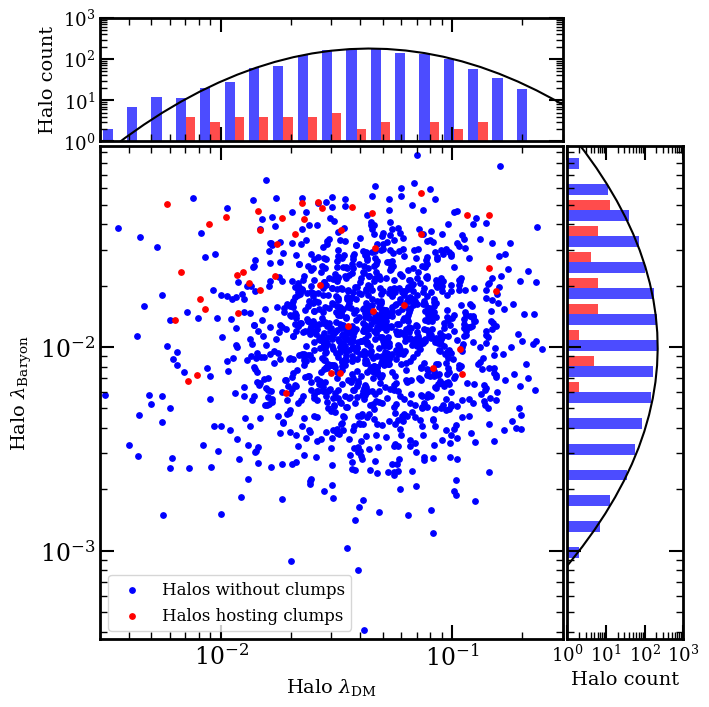}

   \caption{Plot of the spin parameter for the DM versus the baryonic one  at $z = 18$. The red points are the halos hosting clump. On top and on the side are the distribution for the two quantities, again red are halos with clump and blue the all distribution of halos. The histograms are fitted with a lognormal distribution (black curve), with the parameters $\left( \overline{\lambda_{\mathrm{DM}}}, \sigma_{\lambda, \mathrm{DM}} \right) = \left( 0.055, 0.76 \right)$ and $ \left( \overline{\lambda_{\mathrm{baryon}}}, \sigma_{\lambda, \mathrm{baryon}} \right) = \left( 0.014, 0.75 \right)$.
      \label{fig:spin_parameter_distribution}}
\end{figure}

Figure \ref{fig:spin_parameter_distribution} shows the correlation between the spin parameter from the DM and that of the baryons in our simulation at $z = 18$. The red points are the halos hosting gas clump and the blue one show the ones without. As shown by the two histograms, the distribution of the spin parameter follows a lognormal distribution:

\begin{equation}
   P\left(\lambda^{\prime}\right)=\frac{1}{\lambda^{\prime} \sqrt{2 \pi} \sigma_{\lambda}} \exp \left(-\frac{\ln ^2\left( \lambda^{\prime} / \overline{\lambda} \right)}{2 \sigma_{\lambda}^2}\right)
.\end{equation}

We find $\left( \overline{\lambda_{\mathrm{DM}}}, \sigma_{\lambda, \mathrm{DM}} \right) = \left( 0.055, 0.76 \right)$ and $ \left( \overline{\lambda_{\mathrm{baryon}}}, \sigma_{\lambda, \mathrm{baryon}} \right) = \left( 0.014, 0.75 \right)$. The fitted value for the DM is in agreement with \cite{sasakiStatisticalPropertiesDark2014} and \cite{yoshidaEarlyStructureFormation2003}. Furthermore, \cite{sasakiStatisticalPropertiesDark2014} performed high-resolution DM-only simulations with the code \gadget{} to study the properties of DM halos. The spin parameter is of the same order of magnitude for the two components and we observe a larger scatter in the spin of the baryons than for the DM.
There is no correlation between the two components. \cite{desouzaDarkMatterHalo2013} also found a much weaker correlation at this redshift compare to lower redshift as the two components have more time to interact and redistribute angular momentum at lower redshift. From the probability distribution function (PDF) of the two components, we see that halos hosting clumps tend to have a lower DM spin parameter and a higher baryonic one.

\begin{figure}
   \centering{}
   \includegraphics[width=0.9\linewidth]{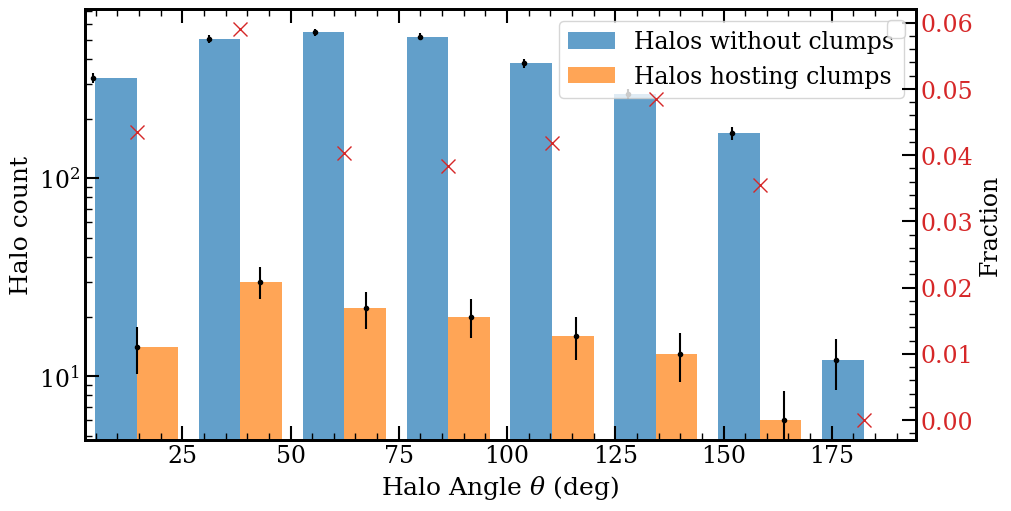}

   \caption{Histogram of the probability density of the angle $\theta$ between the DM and baryonic angular momentum vectors for halos without clumps (blue bars) and halos hosting clumps (red bars) for the three last snapshots (corresponding to redshifts 19.5, 18.8 and 18). Error bars correspond to the Poissonian error. The fraction on the right-axis corresponds to the ratio of the number of halos hosting a clump versus the total number of halos.
      \label{fig:angle_parameter}}
\end{figure}

The angle between the angular momentum of the DM and the baryons, $\theta$, can reveal the degree of interaction between the two components. Figure \ref{fig:angle_parameter} shows the distribution of $\theta$ for all halos and those hosting clumps for the three last snapshots (at $z = $ 19.5, 18.8 and 18) \footnote{We stack three snapshots here to reduce noise due to few halos hosting clumps}.
The distribution shows a trend with a peak at $40^{\circ}$ and a slow decrease toward larger angle. The average value is roughly $70^{\circ}$ for all halos and for halos hosting clumps, meaning that the DM and baryons velocity fields are not aligned. The distribution of the angle is the same for the two populations of halos.

\subsection{Shape and triaxiality parameter}

We are also interested in the shape of the detected halos and clumps. To do so, we compute the triaxiality from the gas and the DM component. 
Indeed, the free-fall time of a collapsing object depends on its aspect ratio \citep{ponAspectRatioDependence2012}, thus on its triaxiality. The triaxiality also gives a rough estimate of the shape, mainly the oblateness and prolateness of a clump.

\subsubsection{Halo virial scale}

To compute the triaxiality, we need to find the eigenvalues of the inertia tensor $I$. We compute the inertia tensor with the classic definition, as used by \cite{desouzaDarkMatterHalo2013},

\begin{equation}
   I_{j k}=\sum_{i=1}^N m_i\left(r_i^2 \delta_{j k}-r_{i j} r_{i k}\right),
\end{equation}
where $r_i$ and $m_i$ are the distant with respect to the halo centre and the mass of a DM matter or a gas cell, while $\delta_{j k}$ is the Kronecker delta. The sum is computed over all cells and particles of a given halos,  $N$ is the number of elements.\\

The eigenvalue value from $I$, $a \geq b \geq c$ correspond to the axis ratio of an ellipsoid representing the equivalent homogenous shape of the halo. The axis ratio can be normalize, to account for the sphericity, $s = c/a$ (a spherical halo has $s = 1$), the oblateness, $q = b/a$, and the prolateness, $p = c/b$. From this definition, the triaxiality is:

\begin{equation}
   T=\frac{1-q^2}{1-s^2}=\frac{a^2-b^2}{a^2-c^2}
,\end{equation}
if $q=1$ (meaning $a=b \geq c$), $T\rightarrow 0$ and the halo is shaped like an oblate (disk-shaped), whereas if $p=1$ (meaning $a \geq c=b$), $T\rightarrow 1$ and the halo is shaped like a prolate (filament-shaped).

\begin{figure}
   \centering{}
   \includegraphics[width=1\linewidth]{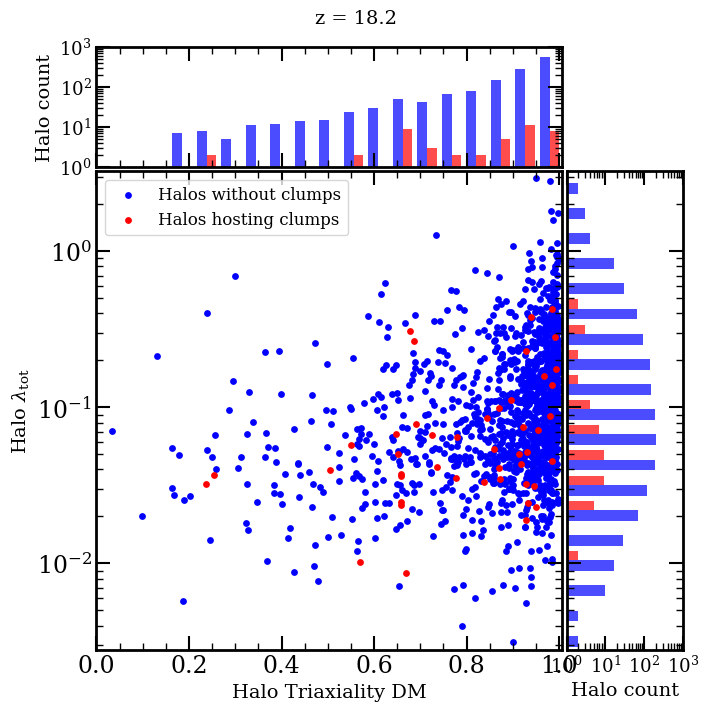}

   \caption{Plot of the spin parameter versus the triaxiality parameter in our simulation, at $z = 18$ for the halos.
      \label{fig:spin_vs_T_dm_parameter_distribution}}
\end{figure}

We computed the triaxiality parameter for the DM component of every halo. Figure \ref{fig:spin_vs_T_dm_parameter_distribution} shows a plot of the halo DM spin parameter versus the triaxiality parameter with an histogram for both quantity. The distribution of the triaxiality parameter shows that most halos are prolate, namely, they have a filamentary shape. This is coherent with theories predicting halos to accrete matter from filaments. There is a trend linking a higher triaxiality parameter to a higher spin parameter.

\subsubsection{Cold and dense gas scale}

We go on to investigate the properties of the gas at much smaller scales. Our simulation is able to resolve the gas, until $n_{\mathrm{H}} \sim 10^5 \mathrm{cm}^{-3}$, where the gas is unstable and collapsing. It is interesting to study the local properties of the dense gas, as these will be the initial conditions for future zoom-in and isolated high-resolution simulations.



\begin{figure}
   \centering{}
   \includegraphics[width=0.8\linewidth]{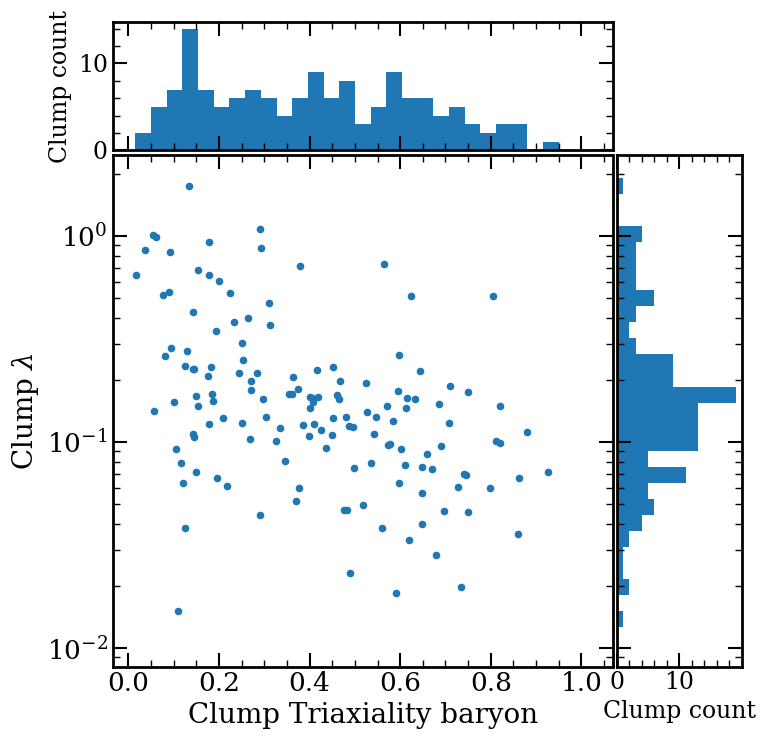}

   \caption{Plot of the spin parameter versus the triaxiality parameter for the gas clouds, at $z = 18$. 
      \label{fig:T_baryon_vs_spin}}
\end{figure}

We computed the spin and the triaxiality parameter for the dense gas clouds following the previous formulation. Here, we computed it only for the gas denser than 100 cm$^{-3}$ to focus on the densest phase within a clump. This threshold corresponds to the point where molecular cooling is the most efficient; thus, the lowest temperature is achieved, which can be seen on the phase diagram. The distribution of the triaxiality parameter is shown in Figure \ref{fig:T_baryon_vs_spin}. It is different from that of the halo distribution (see Figure \ref{fig:spin_vs_T_dm_parameter_distribution}). Here, the distribution is much more uniform. 

The distribution of the triaxiality parameter is nearly uniform between 0 and 1, contrary to the distribution at the halo scale. The presence of oblate geometry of the gas at higher density indicates that some clumps are rotationally supported. \cite{gaoFirstGenerationStars2007} also found such a trend in their simulation. They performed high-resolution cosmological simulations to study the birth site of primordial stars. We observe a slight trend between a larger spin parameter and a larger triaxiality. A higher triaxiality parameter means a more pronounced disk shape, which is expected to be more rotationally supported and thus have a larger spin parameter. We find no correlation between the spin of the dense gas and the one computed at the virial scale from the baryons. The large-scale spin does  not seem to be connected with the dynamic of the gas once it has cooled. 

\subsection{Cooling species}


\label{section:cooling_specie}

Our chemical network includes two cooling species, HD and H$_2$. As H$_2$ is much more abundant than HD, the cooling is mainly driven by H$_2$ in most cases. However, previous studies \citep{ripamontiRoleHDCooling2007,greifSimulationsMovingMesh2011} have identified cases where HD cooling is effective; HD can further cool the gas, even to the CMB floor. This HD-cooling mode has also been identified by \cite{hiranoPrimordialStarFormation2015} in their 3D hydrodynamic simulations. Following this study and that of \cite{mcgreerImpactHDCooling2008}, we can split the halos into HD-cooling ones and others. {We chose to differentiate between these two types of halos by $f_{\mathrm{HD}}/f_{\mathrm{H_{2}}} \geq 10^{-3}$, where $f_{\mathrm{HD}}$ (resp. $f_{\mathrm{H_2}}$) is the HD (resp. H$_2$) mass-fraction of the halo computed for the gas inside its virial radius. We only considered halos with a clump for this analysis as they are the ones directly related to this kind of HD cooling. Of the 73 halos hosting clumps detected at $z = 18$, 36 are HD-cooling, meaning nearly half of them host a high HD fraction. We observe that this ratio tends to increase with redshift, starting from roughly 30\% at $z = 22$ to 50\% at $z=18$. We observed a slightly higher fraction compared with the 10\% fraction found in \cite{hiranoPrimordialStarFormation2015}. We think this might be due to the selection criteria, as we restricted our sample to halos where we detected a clump. Indeed, we have a fraction of 10-12 \% of HD-cooling halos when we consider all halos more massive than $4\times 10^5 \mathrm{M_{\odot}}$.}

\begin{figure}
   \centering{}
   \includegraphics[width=1\linewidth]{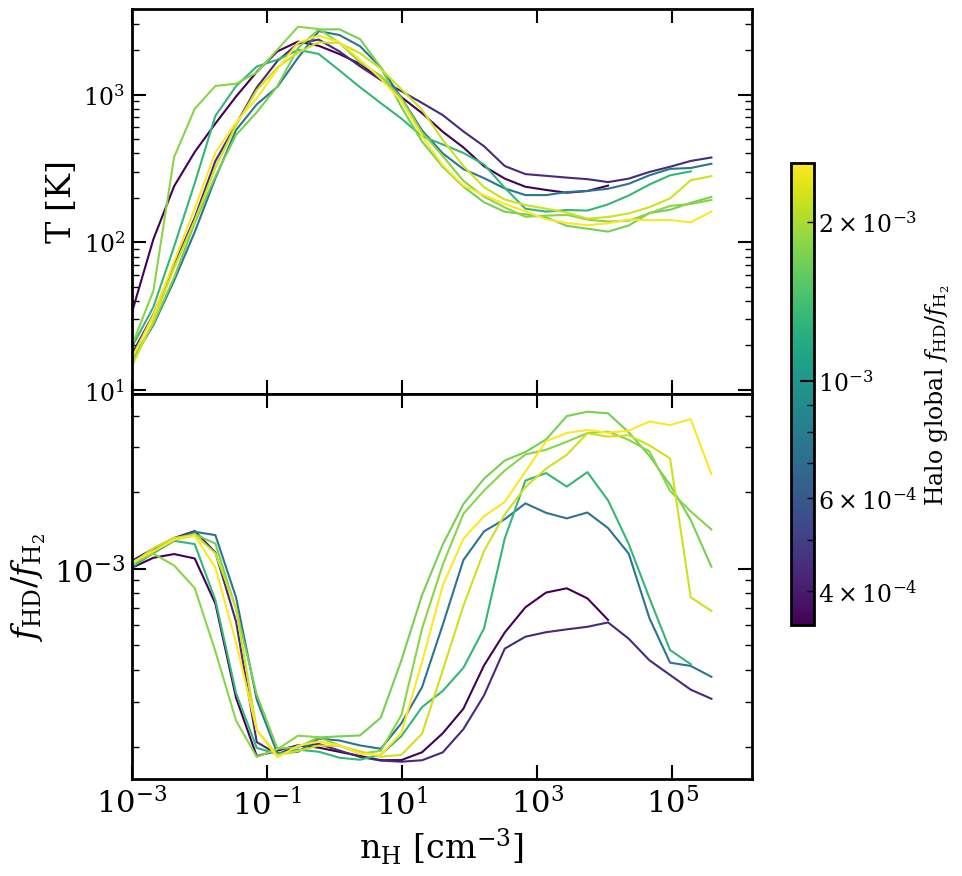}
   \caption{Average profile of primordial halos at $z = 18$. The top plot shows the mass-average gas temperature versus the density, the bottom gives $f_{\mathrm{HD}}/f_{\mathrm{H_{2}}}$ versus the gas density. The color map corresponds to $f_{\mathrm{HD}}/f_{\mathrm{H_{2}}}$. All halos have similar mass, between $1.4 \times 10^6 \mathrm{M_{\odot}}$ and $1.6 \times 10^6 \mathrm{M_{\odot}}$. The only difference among theses is $f_{\mathrm{HD}}/f_{\mathrm{H_{2}}}$, which is indicated via the line color and the color bar on the right. $f_{\mathrm{HD}}/f_{\mathrm{H_{2}}}$ is computed via a mass-average mean of the gas within the virial radius.
      \label{fig:nH_vs_Tgas_ratio_HD}}
\end{figure}

A higher HD fraction is linked with a lower gas temperature among our halo sample. Figure \ref{fig:nH_vs_Tgas_ratio_HD} shows the average profile from various halos with different $f_{\mathrm{HD}}/f_{\mathrm{H_{2}}}$. All halos showed on this plot have similar mass, between $1.4 \times 10^6 \, \mathrm{M_{\odot}}$ and $1.6 \times 10^6 \, \mathrm{M_{\odot}}$. However, their ratio $f_{\mathrm{HD}}/f_{\mathrm{H_{2}}}$ differs at the virial scale, as indicated by the color bar on the right. Then, $f_{\mathrm{HD}}/f_{\mathrm{H_{2}}}$ is between $4 \times 10^{-4}$ and $2.2 \times 10^{-3}$, meaning that some halos are HD-cooling ones. 
The bottom plot from Figure \ref{fig:nH_vs_Tgas_ratio_HD} clearly shows that in the HD-cooling halos, HD formation is only triggered at high density, $n_{\mathrm{H}} \geq 10^2 \, \mathrm{cm^{-3}}$, ie at low temperature $T_{\mathrm{gas}} \sim 1-5 \times 10^2$ K (upper panel), which is what we expect from theory. The upper panel in Figure \ref{fig:nH_vs_Tgas_ratio_HD} clearly shows that HD-cooling halos are colder, nearly reaching $10^2$ K at $n_{\mathrm{H}} = 10^4 \mathrm{cm^{-3}}$, which is still above the CMB cooling limit at this redshift ($\sim$ 50K). We see that HD cooling is non-negligible in the range of $10^2 - 10^6 \, \mathrm{cm^{-3}}$ in our simulation, which is in agreement with \cite{mcgreerImpactHDCooling2008}. Overall, these results are consistent with those from \cite{hiranoPrimordialStarFormation2015,greifSimulationsMovingMesh2011}.

\begin{figure}
   \centering{}
   \includegraphics[width=1\linewidth]{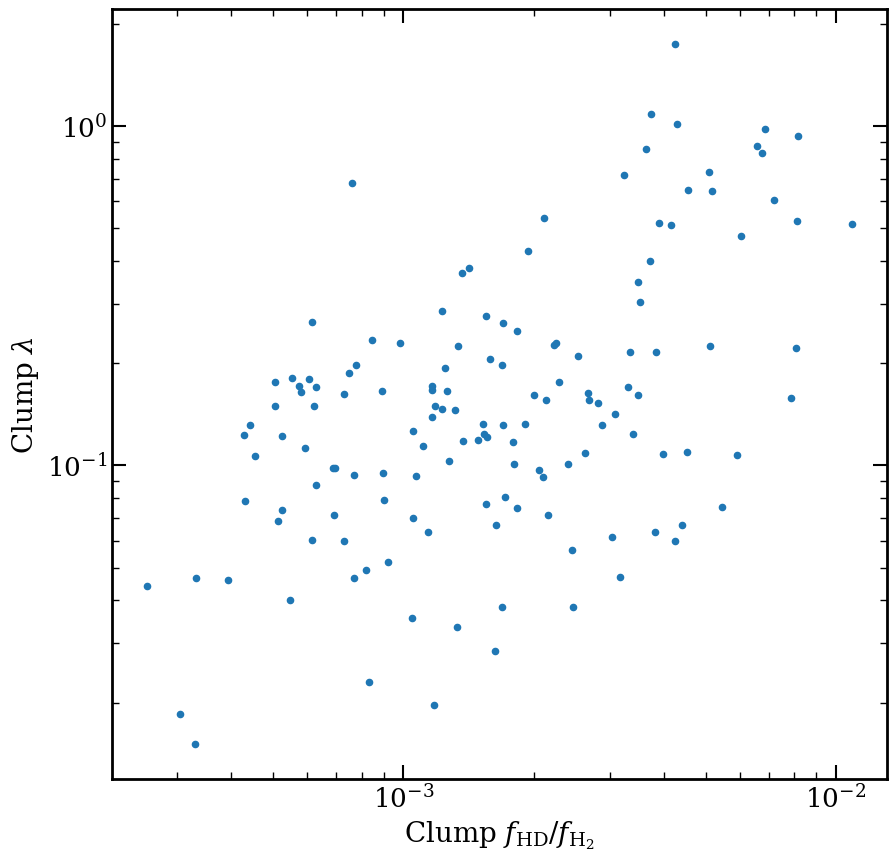}

   \caption{Plot from the mass-average $f_{\mathrm{HD}}/f_{\mathrm{H_{2}}}$ versus the specific spin parameter for each clump of the simulation at $z = 18$. The color coding is the same as before and depends on the threshold. 
      \label{fig:ratio_HD_vs_specific_J}}
\end{figure}

Small differences at the halo scale can be amplified in the collapse and lead to different properties of the star-forming region, in particular the HD fraction. In our simulation, it is not clear what is causing a higher HD fraction in some halos. At the clump scale, we find that a higher $f_{\mathrm{HD}}/f_{\mathrm{H_{2}}}$ is correlated with a lower spin parameter. Figure \ref{fig:ratio_HD_vs_specific_J} shows a plot of the mass-average $f_{\mathrm{HD}}/f_{\mathrm{H_{2}}}$ versus the spin parameter for each clump of the simulation at $z = 18$. We observe a trend linking a higher spin parameter and a higher HD fraction within a clump. One explanation for this behaviour is that the collapse is slower for a rapid rotating cloud which could lead to an enhanced HD formation and more cooling. With a prolonged collapse, there is an extended window for molecular cooling since the creation of molecules does not happen instantly. Furthermore, HD formation is enhanced at low temperature, hence, there is a higher $f_{\mathrm{HD}}/f_{\mathrm{H_{2}}}$ ratio for slow collapse. This  mechanism has been observed by \cite{nishijimaLowmassPopIII2023} and could be efficient here.

\subsection{Spatial distribution of halos hosting clumps}


We have seen that halos hosting star formation have particular physical and chemical properties, in particular, there is a clear mass threshold for a halo to host a cold gas clump. The search for population III birth sites is a search for the most massive halos. However, the properties of halos can be influenced by their environment. For example, \cite{gaoAgeDependenceHalo2005} found that halos assembled earlier are much more clustered and \cite{sasakiStatisticalPropertiesDark2014} found that clustered halos have a higher spin parameters due to stronger tidal forces. To study this point, we computed the two-point correlation function as it highlights such clustering effects. We compared the correlation function for halos with clumps to that of all halos with masses above our resolution limit. We followed the method from \cite{hamiltonBetterWaysMeasure1993}, who computed the correlation function $\xi(r)$ from two catalogues, with the first following a random uniform distribution and that of the simulation. With this, $\xi(r)$ is expressed as:

\begin{equation}
   \xi (r) = \frac{DD(r)\times RR(r)}{RD(r)^{2}} - 1
,\end{equation}
where $DD(r)$, $RR(r),$ and $DR(r)$ are the number of pairs separated by a distance, $r,$ respectively in our simulation (halo-halo pairs), in the random distribution (random-random pairs) and between the random and simulated catalogue (random-halo pairs). 

\begin{figure}
   \centering{} 
   \includegraphics[width=1\linewidth]{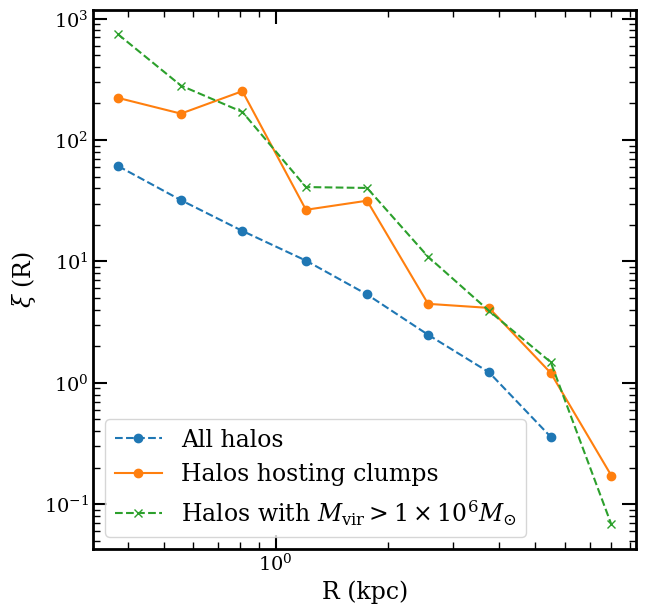}

   \caption{Two-point correlation function of halos in our two simulation at $z = 18$. The dashed blue line shows that of all halos detected, the orange solid line with circles gives that of halos hosting clumps, and the green dashed line with crosses gives that of halos with  $M_{\mathrm{vir}} \geq 1 \times 10^6 M_{\odot}$.
      \label{fig:correlation_fonction}} 
\end{figure}

Figure \ref{fig:correlation_fonction} shows the two-point correlation function of three categories of halos: all halos (blue dotted line), halos hosting clumps (orange solid line), and halos more massive than $M_{\mathrm{crit}} = 1 \times 10^6 M_{\odot}$. Halos hosting clumps are as much clustered as the ones more massive than $1 \times 10^6 M_{\odot}$. The mean separation between two halos hosting clumps is only 0.5 pkpc. 
The correlation function of halos hosting star formation is coherent with that of halos more massive than $1 \times 10^6 M_{\odot}$, the threshold identified previously. The presence of a clump is not influenced by the clustering of a halo. Then, as massive halos are more clustered than all halos, we observe the same behaviour for halos with clump. 

\section{Discussion}
\label{section:discussion}

\subsection{Birth sites of primordial stars}


Our cosmological simulation enables us to probe the condition in which primordial stars were born. The minimum mass for halos hosting star formation we derive in our simulation is consistent with previous results. The minimum mass is primarily set by the fraction of molecular hydrogen produced in the DM halos, which is linked to its virial temperature. Because of the limited volume of our simulation (1 h$^{-1}$cMpc), we are not able to capture the very first star-forming halos, appearing in rare density fluctuations ($\sigma \geq 4$) of the gaussian density field. They were probably born at much higher redshift ($z \geq 30$)  than this study can capture. However, they are extremely rare and so, they are not representative of the global primordial halo population. In order to resolve the birth of the first primordial stars, we would need either a much larger box size or to  artificially simulate an over-dense region of the Universe. 

The chemical properties of halos hosting collapsing gas clumps are quite universal concerning molecular hydrogen. In all of those halos, we observe a similar H$_2$ fraction. However, we observe differences in the HD fraction, which are linked with a higher degree of rotation at the star-forming cloud scale.

\subsection{Caveats}

In our simulations, we did not include streaming velocities between baryons and DM. The relative velocity decreases with the redshift as $v_{\mathrm{stream}} \propto (1+z)$. \cite{druschkeShapeSpinMinihaloes2020} found that streaming velocity has a negligible effect on the spin and shape of the DM component in mini-halos, but strongly affects the gas component. Its spin parameter increases and it is less spherical. Simulations studying the link with primordial star formation shows that it delays the formation of the first pop III stars. \cite{schauerInfluenceStreamingVelocities2019} study this effect in a cosmological simulation. There results without streaming velocity are in agreement with our work, but when they include non-zero streaming velocity the minimum mass for halo formation is increased by a factor of a few. In the most extreme case, star formation occurs exclusively in atomic cooling halo, with $T_{\mathrm{vir}} \geq 10^4$ K. This point is left for future studies. 

We also do not include magnetic fields in our cosmological simulation. The mean magnetic field of the early Universe is not constrained. Even the presence of any magnetic field before the first stars is in question \citep{attiaCosmologicalMagnetogenesisBiermann2021}. \cite{stacyMagneticFieldsFormation2022} compute the primordial IMF with magnetic field. They start from a uniform field $B_0 = 4.5 \times 10^{-12}$ G at $z$ = 54. The growth of the field is then nearly proportional to $n^{2/3}$ in the range $10^{-2} - 10^{8} \, \mathrm{cm^{-3}}$, reaching $10^{-2} $ G at $n_{\mathrm{H}} = 10^8 \mathrm{cm^{-3}}$. The effects of magnetic field are negligible is the range of our present simulation. However, \cite{stacyMagneticFieldsFormation2022} found that is has a significant influence of the primordial IMF by suppressing fragmentation and, thus, low-mass stars as well. The primordial magnetic field at the halo scale probably requires additional study and further constraints to get coherent seeds at the halo scale. 

The first feedback of pop III stars is radiative feedback from the photons emitted by the star. These photons, in particular those in the Lyman-Werner (LW) bands, can photo-ionize and photodissociate molecular hydrogen in a nearby halo. The change in the chemical composition can influence the result of the collapse, to form the so called population III.2 stars \citep{greifFirstGalaxiesAssembly2008,hiranoONEHUNDREDFIRST2014}: a distinct population of metal-free stars formed in halos illuminated by a neighbouring already formed pop III star. Such background radiation can increase the ionisation fraction and photo-dissociate H$_2$. A higher ionisation fraction acts as a catatalyst for H$_2$ formation but Lyman-Werner photons can also dissociate H$_2$. To study this first feedback, we need to estimate the spatial distribution of halos forming stars. We have seen, in a qualitative sense, that halos hosting star formation are as clustered as other halos, highlighting the role of radiative feedback. The timescale for mechanical and chemical feedback is much longer than the radiative one, but plays a major role in the transition between population III and II stars.


\section{Conclusion} 
\label{section:conclusions}

In this study, we run a cosmological simulation following the evolution of  DM and baryons to study the birth sites of primordial stars. Our study makes it possible to obtain the statistical properties of the low-mass halos hosting primordial star formation. The resolution of our cosmological simulation enables us to resolve the star forming region until sufficiently high density $\sim 10^5 \mathrm{cm^{-3}}$ to ensure the gravitational instability of the cold and dense gas cloud. We use an updated chemical network following the abundances of keys species (e$^{-}$, H, H$^+$, H$^-$, D, D$^+$, He, H$_2$, H$_2^+$, HD, and HeH$^+$). We now plan to run zoom-in simulations and isolated collapse to sample the IMF with sink particles.

Our analysis is focused on many important properties of the halos, in particular, the distribution of mass, spin, shape, and the chemical properties. We also briefly study the properties of the dense clouds. Our main results can be summarised as follows : 

$\bullet$ The gas radial profiles are coherent with the global physical understanding of the accretion mechanism. We observe the formation of a shock at small scales close to the centre of the halo, which heats up the gas to $10^4$ K at $n_{\mathrm{H}} = 10^1 \mathrm{cm}^{-3}$. 

$\bullet$ The minimal halo mass for halo hosting pop III star formation is found to be $\sim 7 \times 10^5 M_{\odot}$, in coherence with previous results. We observe a resolution convergence for this lower limit. We also observe a minimum H$_2$ abundance for a halo of $x_{\mathrm{H_2}} \simeq 2 \times 10^{-4}$ at the halo scale. 

$\bullet$ Halo history does not have a major impact on clump formation. However, a high accretion rate can slightly delay the presence of star-forming clump. 


$\bullet$ The spin parameter of the DM is independent of that of baryons at the halo scale. We also observed no correlation between the direction of the angular momentum for the two components. 

$\bullet$  At the halo scale, halos tend to have filamentary shapes. At the clump scale, the distribution of the gas triaxiality parameter is flat. It means that the disk shape is more frequent, which can be due to a larger degree of rotation. 

$\bullet$ The HD abundance is not uniform among the halo population. Halos with a higher HD abundance are colder as the temperature in the range $10^2 - 10^4 \, \mathrm{cm^{-3}}$ depends a lot on the HD abundance. The higher fraction of HD is linked with a higher spin parameter of the dense gas.

\section*{Acknowledgements}
We thanks Maxime Gontel, Pierre Hily-Blant and Alexandre Faure for their help and for letting us use their chemical network. We gratefully acknowledge support from the PSMN (Pôle Scientifique de Modélisation Numérique) of the ENS de Lyon for the computing resources. The authors are grateful to the LABEX Lyon Institute of Origins (ANR-10-LABX-0066) Lyon for its financial support within the Plan France 2030 of the French government operated by the National Research Agency (ANR). The authors are grateful to the LABEX Lyon Institute of Origins (ANR-10-LABX-0066) Lyon for its financial support within the program "Investissements d'Avenir" of the French government operated by the National Research Agency (ANR).
 
\bibliography{Zotero_library.bib, local_lib.bib}

\begin{appendix}

\section{Halo mass function and the higher resolution simulation}

\label{appendix:high_res_simulation}

Reproducing the halo mass function from numerical simulations is a real challenge and suffers from a great deal of caveats, especially at high redshift. One difficulty is reaching a high mass resolution in a volume large enough to sample the cosmological mass perturbation spectrum completely. At high redshifts, the effects of the finite box size become particularly important because any overdensities represent rare fluctuations in the spectrum of linear fluctuations (\cite{reedHaloMassFunction2007}). Also, the box size cuts all larger fluctuations than those that can fit within it. The modes with a large wavelength are thus not present. Another effect linked with the finite box size is the cosmic variance, namely, the run-to-run variations introduced by the finite sampling of density modes. As stated in \cite{lukicHaloMassFunction2007}, the variance within a same set of simulations can induce variation as much as an order of magnitude. Finally, The  parameters of the halo finder can also induce variations. We check (in Section \ref{subsection:halo_MF}) how our fiducial simulation reproduces the halo mass function.

\begin{figure}
   \includegraphics[width=1\linewidth]{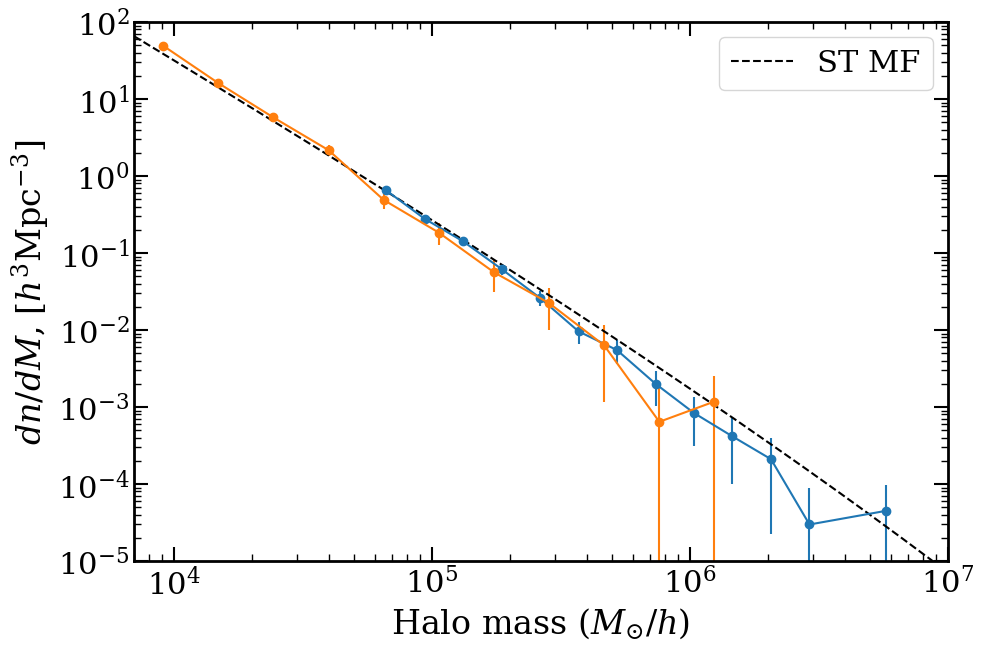}
   \caption{Comparison of the Sheth-Tormen mass function (dashed line) and the one detected in our two simulations at $z = 18$ (solid line with circle). The error bars are the Poissonian error bars. 
   \label{fig:HMF_comparison_2_simu}
   }
\end{figure}

Furthermore, we tested the resolution convergence of this simulation with a higher resolution simulation with the parameter stated in Table \ref{table:parameter_simulation}. We plot the halo mass function of the two simulations on Figure \ref{fig:HMF_comparison_2_simu} (same format as Figure \ref{fig:ST_VS_simulation}). The sample of the halo mass function is valid until $10^4 M_{\odot}$ in the high-resolution simulation, which further extends our resolution limit. 

\begin{figure}
   \includegraphics[width=1\linewidth]{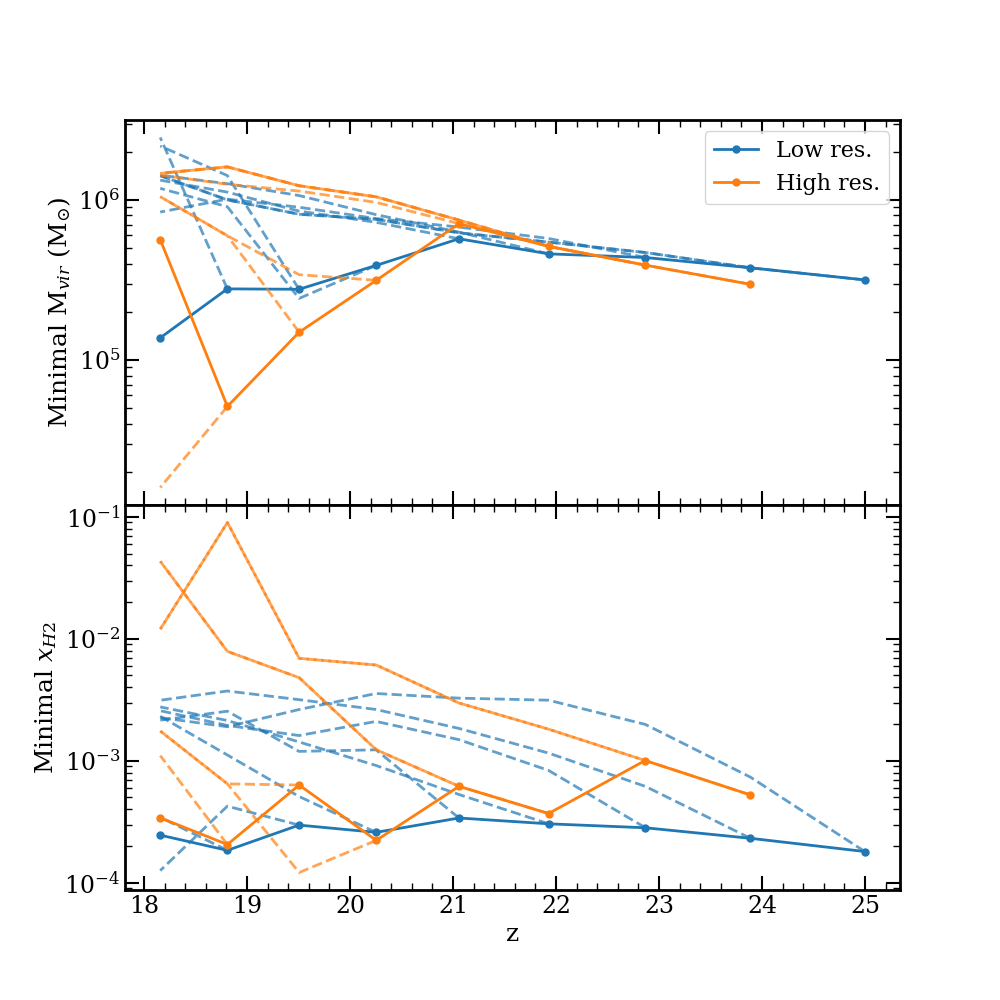}
   \caption{Minimum H$_2$ mass fraction of halos hosting gas clumps detected with our algorithm. The blue (resp. the orange) line corresponds to the high (resp. the low) resolution simulation. 
   The dashed lines show the subsequent mass or $x_{\mathrm{H_2}}$ evolution of the relevant 'minimal' halo, usually increasing in both mass and H2 fraction with the same color-coding.
   \label{fig:Min_H2_mass_halo_HR}
   }
\end{figure}

To test the convergence, we reproduce Figure \ref{fig:Min_H2_mass_halo} which shows the minimal mass and H$_2$ fraction of halos hosting cold gas clump. The plot corresponding to the two simulation are shown on Figure \ref{fig:Min_H2_mass_halo_HR}. In these two panels, the blue curve is the same as that of Figure \ref{fig:Min_H2_mass_halo_HR}. The curve from the high-resolution simulation (the orange line) is in agreement with the blue one, showing that the two simulations agree for the low-mass halo limit. 

\section{Chemical initial conditions}
\label{appendix:chem_initial_cdt}

We computed the chemical abundances at $z=100$, thanks to a one-zone model that takes into account the physical evolution of the Universe from $z=10^{4}$. Our network includes photo-ionisation and photo-dissociation by the CMB photons and it is still valid at higher redshift. Computing the initial chemical abundances in this way enables us to be consistent with our \ramses{}-\krome{} simulation, starting at redshift $z=100$.

After the primordial nucleosynthesis, the Universe only consists of
hydrogen, deuterium, helium and a tiny fraction of lithium. At a redshift of
$z=10^{4}$, all atoms are fully ionised as the thermal radiation
is above there excitations levels \citep{seagerHowExactlyDid2000}. The
recombination starts at a redshift of 1100 and at $z=100$, the Universe
is nearly neutral. However, the electron fraction and the number density
of the molecules is not zero, even at this redshift. To study
the formation of the first luminous objects, we need to know the initial
chemical abundances of every species. This is of primordial importance
as molecular formation is catalysed and strongly enhanced by the ionisation fraction, for
example. 

We considered a one-zone spherical Universe. The density is assumed
to be uniform at all times and it corresponds to the co-moving density
: $\rho=\Omega_{\mathrm{B}}\rho_{c}a^{3}$, where $\Omega_{\mathrm{B}}$ is the density 
parameter, $\rho_{c}$ is the critical density, and $a$ is the expansion
factor. We set the radiation temperature as: $T_{\mathrm{rad}}=T_{\mathrm{CMB}} \times a$,
with $T_{\mathrm{CMB}}=2.73$ K (the temperature of the CMB). We started at $z=10^{4}$,
assuming that all species are fully ionised, taking $\frac{n_{He}}{n_{\mathrm{H}}}=0.082$
and $\frac{n_{D}}{n_{\mathrm{H}}}=4.65\times10^{-5}$. We chose to start the one-zone simulation at $z=10^{4}$ with all species fully ionised.
We took the maximum ionised level for all species: H$^{+}$, D$^{+}$,
He. Our network does not include He$^{++}$ and He$^{+}$, so we started our simulation with only He, which is not realistic at high redshifts, but this choice does not influence the values found at $z=100$.

\begin{figure}
   \centering{}
   
   \includegraphics[width=1\linewidth]{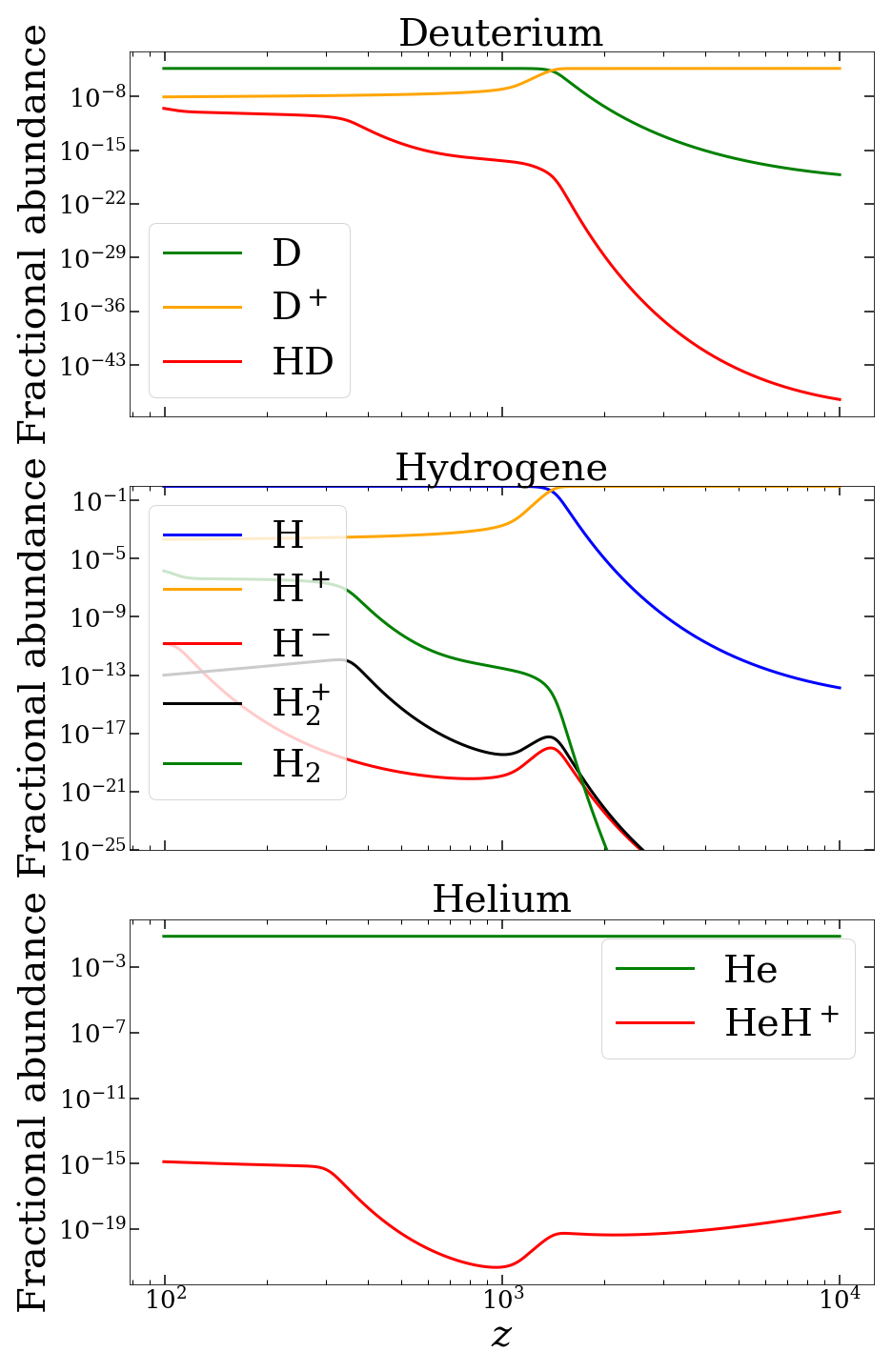}

   \caption{Evolution of the species as a function of redshift. Top : Deuterium species (D, D$^+$ and HD), Middle : Hydrogen species (H, H$^+$, H$^-$, H$_2^+$ and H$_2$), Bottom : Helium species (He and HEH$^+$)\label{fig_chemical_abundance}}
\end{figure}

Figure \ref{fig_chemical_abundance} shows the evolution of abundances of different species of the early Universe. Our results are consistent with simulation of molecule formation in the early Universe \citep{novosyadlyjMoleculesEarlyUniverse2017} and from other primordial networks. 

\begin{table}
   \caption{Chemical abundances at the redshifts, $z=10^{4}$, $z=10^{3}$, and $z=100$}             
   \label{table:abundance_chemical}      
   \centering                          
   \begin{tabular}{c c c c}        
   \hline\hline                 
   Species & $z=10^{4}$ & $z=10^{3}$ & $z=100$ \\    
   \hline                        
   E        &  0.924                       &  1.894 $\times 10^{-3}$    &  2.106 $\times 10^{-4}$   \\
   H-       &  0                          &  1.243 $\times 10^{-20}$      &  1.406 $\times 10^{-11}$  \\
   H        &  0                          &  0.922                          &  0.923                       \\
   H2       &  0                             &  2.911 $\times 10^{-13}$   &  1.434 $\times 10^{-06}$  \\
   D        &  0                             &  4.29 $\times 10^{-05}$    &  4.298 $\times 10^{-05}$  \\
   HD       &  0                             &  4.055 $\times 10^{-17}$   &  2.696 $\times 10^{-10}$  \\
   HE       &  0.076                         &  0.076                       &  0.076                       \\
   H$^+$    &  0.924                         &  1.894 $\times 10^{-3}$    &  2.106 $\times 10^{-4}$   \\
   H2$^+$   &  0                             &  3.545 $\times 10^{-19}$   &  1.001 $\times 10^{-13}$  \\
   D$^+$    &  4.29 $\times 10^{-05}$   &  8.724 $\times 10^{-08}$        &  8.406 $\times 10^{-09}$  \\
   HEH$^+$  &  0                          &  4.896 $\times 10^{-22}$   &  1.315 $\times 10^{-15}$  \\
   
   \hline                                   
   \end{tabular}
   \end{table}

\section{Chemical network}
\label{appendix:chemical_network}

Here, we present the chemical network used in the simulation. Full details about the network, in particular, the rates, will be described in a future paper: \cite{FaureNetwork2023}. The molecular abundances, especially those of H$_2$ and HD, were found to be in good agreement with previous studies, suggesting that the dominant (thermal) reaction rate coefficients have reached such a level of accuracy that they are no longer a limiting factor. An important reaction that needs further investigations is the charge exchange between H$_2^+$ and H.  The full chemical network is available upon request.

\begin{table*}
   \centering
   \begin{tabular}{ c c}
      \hline\hline                 
      Label & Reaction  \\
      \hline
      H1 & $\mathrm{H}^{+}+\mathrm{e}^{-} \rightarrow \mathrm{H}+\nu$ \\
      H2 & $\mathrm{H}+\nu \rightarrow \mathrm{H}^{+}+\mathrm{e}^{-}$ \\
      H3 & $\mathrm{H}^{-}+\nu \rightarrow \mathrm{H}+\mathrm{e}^{-}$ \\
      H4 & $\mathrm{H}+\mathrm{e}^{-} \rightarrow \mathrm{H}^{-}+\nu$ \\
      H5 & $\mathrm{H}^{-}+\mathrm{H} \rightarrow \mathrm{H}_2+\mathrm{e}^{-}$ \\
      H6 & $\mathrm{H}^{-}+\mathrm{H}^{+} \rightarrow \mathrm{H}+\mathrm{H}$ \\
      H7 & $\mathrm{H}+\mathrm{H}^{+} \rightarrow \mathrm{H}_2^{+}+\nu$ \\
      H8 & $\mathrm{H}_2^{+}+\nu \rightarrow \mathrm{H}+\mathrm{H}^{+}$ \\
      H9 & $\mathrm{H}_2^{+}+\mathrm{H} \rightarrow \mathrm{H}_2+\mathrm{H}^{+}$ \\
      H10 & $\mathrm{H}_2+\mathrm{H}^{+} \rightarrow \mathrm{H}_2^{+}+\mathrm{H}$ \\
      H11 & 3 $\mathrm{H} \rightarrow \mathrm{H}_2 + \mathrm{H}$ \\
      H12 & $\mathrm{H}_2 + \mathrm{H} \rightarrow 3 \mathrm{H} $ \\

      \hline
      \rule{0pt}{3ex}    
      D1 & $\mathrm{D}^{+}+\mathrm{e}^{-} \rightarrow \mathrm{D}+\nu$ \\
      D2 & $\mathrm{D}+\nu \rightarrow \mathrm{D}^{+}+\mathrm{e}^{-}$ \\
      D3 & $\mathrm{D}^{+}+\mathrm{H} \rightarrow \mathrm{D}+\mathrm{H}^{+}$ \\
      D4 & $\mathrm{D}+\mathrm{H}^{+} \rightarrow \mathrm{D}^{+}+\mathrm{H}$ \\
      D5 & $\mathrm{D}^{+}+\mathrm{H}_2 \rightarrow \mathrm{H}^{+}+\mathrm{HD}$ \\
      D6 & $\mathrm{HD}+\mathrm{H}^{+} \rightarrow \mathrm{H}_2+\mathrm{D}^{+}$ \\
      \hline
      \rule{0pt}{3ex}
      He1 & $\mathrm{He}^{+}+\mathrm{e}^{-} \rightarrow \mathrm{He}+\nu$ \\
      He2 & $\mathrm{He}+\nu \rightarrow \mathrm{He}^{+}+\mathrm{e}^{-}$ \\
      He3 & $\mathrm{He}+\mathrm{H}^{+} \rightarrow \mathrm{HeH}^{+}+\nu$ \\
      He4 & $\mathrm{HeH}^{+}+\nu \rightarrow \mathrm{He}+\mathrm{H}^{+}$ \\
      He5 & $\mathrm{HeH}^{+}+\mathrm{H} \rightarrow \mathrm{He}+\mathrm{H}_2+$ \\
            \end{tabular}

      \caption{Reaction of the chemical network.}

\end{table*}

\end{appendix}

\end{document}